\documentclass[12pt,a4paper,final]{article}

\usepackage[latin1]{inputenc}
\usepackage[round]{natbib}
\usepackage{amssymb, amsfonts, amsmath}
\usepackage{graphicx}
\usepackage{dcolumn}
\usepackage{eurosym}    
\usepackage[left=1.1in, right=1.1in, top=1in, bottom=1in]{geometry}
\usepackage{setspace}
\usepackage{tabularx}
\usepackage{float}
\usepackage{booktabs}
\usepackage{titling}    
\usepackage{verbatim}   
\usepackage{lscape}     
\usepackage{multirow}
\usepackage{caption}
\usepackage[OT1]{fontenc}
\usepackage[bitstream-charter]{mathdesign}
\usepackage{color}
\usepackage{caption}
\usepackage{subcaption}
\usepackage{lscape}
\usepackage{rotating}
\usepackage{todonotes}
\usepackage{soul}
\usepackage{xcolor}
\usepackage{titlesec} 
\usepackage[hyphens]{url}

\title{\vspace{-1in}Optimal Targeting in Fundraising: \\ A Causal Machine-Learning Approach}

\author{Tobias Cagala, Ulrich Glogowsky, \\ Johannes Rincke, Anthony Strittmatter\thanks{Cagala: Deutsche Bundesbank (tobias.cagala@bundesbank.de); Glogowsky: Johannes Kepler University Linz and CESifo (ulrich.glogowsky@jku.at); Rincke: University of Erlangen-Nuremberg (johannes.rincke@fau.de); Strittmatter: CREST-ENSAE, Institut Polytechnique Paris and CESifo (anthony.strittmatter@ensae.fr). {Acknowledgments:} We have benefited from comments by Maja Adena, Elliott Ash, Pierre Boyer, Bruno Cr\'{e}pon, Philipp D\"{o}rrenberg, Michael Knaus, Emanuel Hansen, Andreas Haufler, Steffen Huck, Michael Lechner, Benjamin M. Marx, Dominik Papies, Michael P. Price, Dominik Sachs, Benoit Schmutz, Stefan Wager, and Zhengyuan Zhou. We also thank conference and seminar participants at the European Meeting of the Econometric Society 2021, the ETH Zurich 2021, the Data Science, Statistics \& Visualisation conference 2021, the Causal Data Science Meeting 2020, the Annual Conference of the National Tax Association 2019, the Annual Conference of the German Economic Association 2019, the Annual Conference of the International Institute of Public Finance 2019, and the WZB Workshop on Charitable Giving in 2018. We gratefully acknowledge financial support from the Swiss National Science Foundation (Spark Project 190422, Strittmatter), the French National Research Agency (LabEx Ecodec/ANR-11-LABX-0047, Strittmatter), the Hans-Frisch-Stiftung, and the Emerging Field Initiative at University of Erlangen-Nuremberg (Cagala, Glogowsky, and Rincke). The paper represents the authors' personal opinions and does not necessarily reflect the views of the Deutsche Bundesbank or its staff. AEA RCT Registration ID: AEARCTR-0007581.}}
\date{\today } 

\def\ci{\perp\!\!\!\perp}

    \newcolumntype{L}[1]{>{\raggedright\arraybackslash}p{#1}}
    \newcolumntype{C}[1]{>{\centering\arraybackslash}p{#1}}
    \newcolumntype{R}[1]{>{\raggedleft\arraybackslash}p{#1}}

    \newcommand{\subtitle}[1]{%
      \posttitle{%
        \par\end{center}
        \begin{center}\Large#1\end{center}
        \vskip0.5em}%
    }

\usepackage[colorlinks=true,linkcolor=blue, citecolor=blue, allcolors=blue]{hyperref}

	\titlespacing*{\paragraph}{0pt}{1.5ex}{1em}
	\titlespacing*{\section}{0pt}{2ex}{0.5em}
	\titlespacing*{\subsection}{0pt}{2ex}{0.5em}

\begin{document}
\maketitle
\thispagestyle{empty}

\begin{center}
	\noindent\textbf{Abstract}\\
\end{center}
\small
Ineffective fundraising lowers the resources charities can use to provide goods. We combine a field experiment and a causal machine-learning approach to increase a charity's fundraising effectiveness. The approach optimally targets a fundraising instrument to individuals whose expected donations exceed solicitation costs. Our results demonstrate that machine-learning-based optimal targeting allows the charity to substantially increase donations net of fundraising costs relative to uniform benchmarks in which either everybody or no one receives the gift. To that end, it (a) should direct its fundraising efforts to a subset of past donors and (b) never address individuals who were previously asked but never donated. Further, we show that the benefits of machine-learning-based optimal targeting even materialize when the charity only exploits publicly available geospatial information or applies the estimated optimal targeting rule to later fundraising campaigns conducted in similar samples. We conclude that charities not engaging in optimal targeting waste significant resources.
\normalsize

\medskip\bigskip
\noindent \textit{JEL codes}: C93; D64; H41; L31; C21

\noindent \textit{Keywords}: charitable giving, optimal policy learning,individualized treatment rules


\section{Introduction}

Fundraising is a costly activity: The 25 largest US charities spend between $5\%$ and $25\%$ of total donations on fundraising expenses \citep{And2011}. These numbers are a matter of concern for two reasons. First, high fundraising costs leave a smaller proportion of overall donations to finance charitable projects. This can lead to an underprovision of the goods and services that charities provide and may, thus, lower welfare if the donors' utility depends on provision levels \citep{Rose1982, Name2013}. Second, high fundraising costs also matter from the charity's perspective because donors are averse to financing overhead costs \citep{Tin2007, Gne2014}. Hence, charities with excessive fundraising expenses will be less successful in raising donations. In conclusion, reducing disproportional fundraising costs can be crucial, from both the welfare and charity-management perspectives. However, while there is broad literature studying how different fundraising instruments such as matching grants and unconditional gifts affect donors' behavior \citep[surveyed by][]{And2013}, previous research has paid less attention to how charities can increase the cost-effectiveness of fundraising.

In this paper, we exploit a causal machine-learning-based approach to maximize a charity's fundraising effectiveness: optimal targeting of fundraising activities to potential donors.%
\footnote{See \cite{ath19b} for a review of causal machine-learning methods.}
Machine-learning-based optimal targeting exploits the possibility that, due to heterogeneity in donors' preferences, the effects of any fundraising campaign are likely heterogeneous across individuals with different observable characteristics.%
\footnote{For example, donation motives (such as altruism, warm glow, and reciprocity) are heterogeneously distributed across individuals \citep{falk2018}. Consequently, responses to fundraising activities that leverage (some of) these motives are likely heterogeneous as well.}
Charities that ignore this heterogeneity may engage in loss-leading fundraising, for example, by directing a costly fundraising instrument to notorious nondonors or donors who, in response to the instrument, do not increase their donations enough to cover the instrument's cost. Against this backdrop, our key contribution is to use machine learning to identify a targeting rule (out of all feasible rules) that is optimal in the sense that it maximizes expected profits by avoiding loss-leading solicitations.%
\footnote{A targeting rule is feasible if it is a deterministic function of the observable characteristics.}
Consequently, charities that engage in this type of optimal targeting can increase net donations raised (i.e., donations net of fundraising costs) and, hence, provide more goods and services.

We demonstrate the potential of machine-learning-based optimal targeting using a common fundraising instrument as an example: small unconditional gifts that accompany a solicitation letter. In theory, such unconditional gifts should increase donations by triggering a reciprocal reaction in recipients \citep{Fal2007}. In practice, the evidence on the effectiveness of unconditional gifts is, however, mixed. Some studies suggest that unconditional gifts are an effective fundraising instrument, while others find that they do not affect average donations or even backfire and lower giving (see the literature section for a discussion). This type of effect heterogeneity might very well operate through individual heterogeneities, such as heterogeneous social preferences. Thus, fundraising gifts offer a promising context to study the benefits of machine-learning-based optimal targeting.

The goal of optimal targeting lies in directing a fundraising instrument (a gift in our example) to a subset of individuals such that the fundraising campaign's expected profits are maximized (i.e., the additional expected net donations collected by the campaign). Yet, who should charities target for that purpose? By definition, the \textit{optimal targeting rule} states that the profit-maximizing targets are the so-called net donors: individuals whose expected additional donation is higher than the marginal fundraising cost. In an ideal world, the charity would know each individual's donation with and without the fundraising instrument (i.e., the gift) and could, thus, determine the set of net donors. In reality, however, this set is unknown to the charity. The reason is that donations under both conditions (gift vs. no gift) are unknown before the campaign. Moreover, even after the campaign, the charity could still only learn a donor's behavior under her assigned condition. These complications motivate our approach to identify optimal targets for fundraising activities.

Our approach to optimal targeting relies on two ingredients. First, it exploits random variation in the assignment of an unconditional gift at the donor level. %
To induce this variation, we teamed up with a charity that sends out solicitation letters to its donor base once a year. In this setting, we randomly assigned almost $20{,}000$ potential donors to a gift treatment (in which individuals received the solicitation letter with a gift) and a control group (in which they received only the letter). Second, as previously highlighted, our approach  relies on machine-learning algorithms. Particularly, we let an algorithm learn the relationship between the individuals' observable characteristics and their expected donation behaviors in the experiment's treatment and control groups (i.e., with and without the gift). From this relationship, we then estimate the (out-of-sample) set of predicted net donors who should be targeted. %
This procedure establishes what we label the \textit{estimated optimal targeting rule}. Importantly, although we cannot trace out the causal effect of the characteristics that drive the heterogeneity, the policy-relevant increase in net donations achieved by the targeting rule is identified.

More specifically, our paper draws on the following machine-learning methods. We implement the optimal-policy-learning algorithm of \cite{ath21}, which extends the empirical welfare-maximization approach of \cite{kit18} by machine learning. In our main specifications, we then estimate optimal targeting rules with the exact policy-learning tree of \cite{zho2019} and show the sensitivity of our results to various alternative estimators (logit, logit lasso, classification and regression trees, and classification forest). Regarding data, we feed our algorithm with information from various sources and of different types, including socioeconomic characteristics, past donation data, and geospatial information. The geospatial information consists of publicly available information from Google Maps on economic and cultural facilities close to the potential donor's place of residence.

Our analysis yields two sets of main results. The first set concerns the warm list (i.e., the sample of previous givers). For this sample, we demonstrate that machine-learning-based optimal targeting substantially boosts the charity's net donations compared to benchmarks where everybody receives the gift (increase relative to this benchmark: $13.8\%$) or no one receives the gift (increase: $14.3\%$). We also show that these positive effects on net donations do not merely reflect pull-forward effects (i.e., shifts of donations from a later year to the experimental year). Next, we illustrate that the estimated optimal targeting rule is stable across two identical consecutive fundraising campaigns conducted in similar samples. To that end, we repeated our field experiment in a consecutive year with a new sample of warm-list individuals. We then highlight that, relative to our benchmarks, the charity could substantially increase net donations in the second year if it distributed gifts according to the optimal targeting rule estimated with first-year data. Another insight of the analysis is that our fundraiser can reap the full benefits of machine-learning-based optimal targeting by relying on data that should be easily accessible by all charities. For the gains of our approach to materialize, knowledge about who is in the warm list and publicly available geospatial information is sufficient. Consequently, we expect our approach to be widely applicable.

The second set of results concerns the cold list (i.e., the sample of previous nondonors). In contrast to the warm list, machine-learning-based optimal targeting in the cold list does not increase net donations compared to the no-gift benchmark. Furthermore, the estimated optimal targeting rule does not broaden the donor base enough to justify the gift's additional fundraising costs. We conclude that, in our context, the charity should not target the gift to cold-list individuals at all. Whereas the details of our findings may well be specific to the setting, a general conclusion is that charities that do not optimally target their fundraising efforts waste significant resources.

The paper is organized as follows. Section \ref{sec:literature} outlines our contributions to the literature, Section \ref{sec:background_and_design} explains the institutional background and design of the experiment, and Section \ref{sec:emprical_strategy} describes our empirical strategy. Section \ref{sec:results} discusses our results, and Section \ref{sec:conclusion} concludes. Online Appendices \ref{app:descriptives}--\ref{app:sensitivity} provide supplementary materials.

\section{Contributions to the Literature}
\label{sec:literature}
In the following, we detail how our study relates and contributes to various literature strands in economics and marketing.

\addvspace{0.1cm}\noindent\textbf{Economics literature.}\hspace{0.2cm}
The first relevant strand of literature, the theoretical fundraising literature in public economics, provides a theoretical underpinning for why fundraising targeting can be beneficial. The argument is as follows. Many charities have properties similar to privately provided public goods \citep{And2013}: contributions are voluntary and the provided goods are nonexcludable and nonrivalrous. In such a context, fundraising tools can counteract free-riding and, hence, underprovision problems \citep[see, e.g.,][]{And1988, Mor2000, Vest2003, And2003, And2013}.%
\footnote{Which tool is suited to increase provision depends on the context. Solicitation letters (perhaps providing return envelopes) counteract the underprovision problem in settings with transaction costs \citep{And2003}. By contrast, leadership gifts oppose the underprovision of threshold public goods \citep{And1988} and public goods under imperfect information \citep{Vest2003}. Lotteries, in contrast, can increase efficiency in standard public goods settings \citep{Mor2000}.}
However, if fundraising is costly and charities must compete for donors, competition can push the costs to such high levels that the total service provision falls \citep{Rose1982, Ald2010, Ald2014}. Targeting of fundraising instruments can then be a tool to maximize donations net of fundraising costs and, thus, provision levels. Specifically, \cite{Name2013} show that a charity's optimal strategy is to target net donors. However, for charities, it is challenging to follow this theoretical rule, as they cannot easily identify these optimal targets. Along these lines, our contribution to this literature is to explore how a data-driven machine-learning approach can help charities predict the set of net donors. Moreover, we study the external validity of estimated optimal targeting rules across fundraising campaigns for the first time.

A second emerging literature strand closely related to our work empirically studies how to target fundraising among heterogeneous donors. We are aware of only two papers that examine this topic. First, \cite{Adena2019} apply a targeting strategy to \emph{matching gifts,} a fundraising tool where funds collected before a campaign top up donations above a threshold. Their main result is that charities can crowd in donations by conditioning these thresholds on past giving behavior (i.e., the thresholds are targeted). Second, \cite{Drou2021} focus on belief-based targeting. They conclude that charities can increase net donations by targeting information treatments to potential donors who hold incorrect (low) beliefs about others' donations. Our paper differs from these studies in two dimensions: First, \cite{Adena2019} and \cite{Drou2021} both build on conceptual considerations to identify dimensions of heterogeneity used for targeting. Following \cite{ath21}, our paper takes a more agnostic and more flexible, data-driven approach. Particularly, our machine-learning algorithms autonomously uncover the most influential dimensions of heterogeneity based on observable donor characteristics. Consequently, the resulting estimated targeting rules can flexibly account for many different sources of heterogeneity (e.g., preferences or income), as long as they correlate with individuals' observable characteristics. Second, instead of considering debiasing or threshold matching, we focus on the unconditional gift as a different fundraising tool.

A third relevant literature strand is the literature on fundraising gifts \citep[see, e.g., the review of][]{List2012}. While \cite{Fal2007} documents that unconditional gifts are a cost-effective tool to increase donations in the warm list, other studies paint a more scattered picture. For example, \cite{lan10} find zero effects of gifts in the warm list, and \cite{ALP2008} conclude that conditional on giving, gifts even lower contributions. \cite{yin2020coins} present similar results.%
\footnote{Thank-you gifts that charities hand out after a donation also seem to lower subsequent donations \citep{NEW2012}. \cite{Eck2018} directly compare unconditional and thank-you gifts and show that donors are twice as likely to give when they receive a high-quality unconditional gift.}
We add to the looming discussion on the causes of effect heterogeneity by showing that individual heterogeneity alone is powerful enough to account for gifts' negative and positive effects.%
\footnote{Of course, differences in the overall setting (such as varying causes of charities) might also explain why different studies come to varying conclusions.}
In particular, in our setting, some groups of potential donors increase donations in response to gifts, and other groups reduce their donations. This finding persists if we restrict our sample to the warm list only. While these findings are insightful in themselves, our contribution to this literature is to demonstrate that charities can effectively exploit the effect heterogeneities to target gifts optimally.

By focusing on effect heterogeneity, we contribute to a fourth literature strand, heterogeneous responses to fundraising that identifies five forms of heterogeneity: (a) characteristics of the charitable organization and the purpose of the charity \citep[e.g.,][]{okt00, vri15}, (b) characteristics of the donors \citep[e.g.,][]{andr03,andr01,raj09,wie13}, (c) donation motives or preferences of donors \citep[e.g.,][]{eck12,har07,KIZ2018}, (d) past donation behavior \citep{schl97,has14}, and (e) crowding out \citep{mee17}. Instead of studying single, selected dimensions of heterogeneity, we combine a range of individual characteristics and past donation behavior. Additionally, our paper adds a rarely used determinant of heterogeneity to the analysis, geospatial characteristics. As this information is easily accessible to charities, it is particularly beneficial for optimal targeting.%
\footnote{\cite{don19} and \cite{glae18,glae20} show that geospatial characteristics are proxies for income and socioeconomic characteristics. One reason is neighborhood segregation \citep[see, e.g.,][]{heb20}.}

\addvspace{0.1cm}\noindent\textbf{Marketing literature.}\hspace{0.2cm}
Our paper also relates to two strands of literature on the targeting of marketing interventions. The first strand applies machine-learning techniques to target marketing in contexts other than charitable giving. For example, \cite{GUE2015} and \cite{ASC2018} study targeting of retention efforts to lower customer churn in telecommunications, professional memberships, and insurance. Moreover, \cite{FON2019}, \cite{ELL2020}, \cite{GUB2020}, and \cite{smi2021} study targeting of recommendations, promotions, coupons, and pricing (mainly in online marketing). These studies not only consider contexts very different from ours, but also employ other methods. Particularly, the papers estimate heterogeneous effects of (randomized) marketing interventions and transform them into a targeting rule by discretizing.%
\footnote{In marketing, this procedure is called ``uplift modeling. ''Economists use the label ``effect-based targeting.''}
Our paper, instead, follows \cite{ath21} and estimates the optimal targeting rule directly (instead of first estimating the effect heterogeneity). To the best of our knowledge, our paper is the first application of this novel, more direct method, explicitly tailored for optimal targeting.%
\footnote{Related machine-learning-based literature profiles or segments customers based on their responses under one condition. For example, \cite{CUI2006} and \cite{KIM2005} show that neural networks are valuable tools to improve targeting in response models. Moreover, \cite{ABE2004} successfully apply reinforcement learning to cross-channel marketing, and \cite{SCH2017} apply a multiarmed bandit to target display advertising.}
The second relevant literature strand is the literature on \textit{charitable marketing} \citep[see, e.g.,][]{WIN2013, KIZ2018}. While this literature has studied targeting in charitable giving, it so far has not used the powerful machine-learning-targeting toolkit. To sum up, the available studies deviate from our paper either because they use machine learning in contexts other than charitable giving or because they study targeting in charitable giving without using causal machine-learning techniques. %

\addvspace{0.1cm}\noindent\textbf{Targeting literature.}\hspace{0.2cm}
Methodologically, we contribute to the small but rapidly growing literature that applies machine-learning methods to target public and private policies \citep[e.g.,][]{and18,hit18,kan13,kna18b,kni19,roc11,kle15}. While these papers consider such contexts as taxation and labor-market programs, our study is the first that applies a fully fledged optimal policy learning algorithm like that of \cite{ath21} to the context of charitable giving. 

\section{Experimental Design and Data}
\label{sec:background_and_design}
Our approach to derive a machine-learning-based optimal targeting rule proceeds in three steps. First, we conduct a field experiment that randomly allocates our fundraising instrument, an unconditional gift. Second, we use machine-learning algorithms to estimate the optimal targeting rule for this gift in a random subsample of the experimental data while retaining the remaining sample. Third, we extrapolate the estimated optimal targeting rule to the retained sample and apply off-policy-learning techniques to assess the estimated rule's out-of-sample performance. While this section details the experimental design and data, Section \ref{sec:emprical_strategy} outlines the machine-learning and off-policy learning approaches.

\subsection{The Natural Field Experiment}
In $2014$, we implemented a natural field experiment in collaboration with a fund\-raiser of the Catholic Church that operates in a German urban area.%
\footnote{In $2015$, we implemented a second experiment with two treatments, an unconditional gift treatment and a gift treatment that framed the gift as a reward for past donations. We use this experiment to (a) examine the stability of the targeting rules and (b) evaluate and compare the effects of the differently framed gifts in a companion paper.}
For decades, this fundraiser has organized a large-scale, annual fundraising campaign: Once a year, it has mailed solicitation letters to all resident church members, irrespective of previous donations. This fund drive aims to finance local church-related projects, such as the renovation of clergy houses, parish centers, or churches. Our experiment exploited this campaign by (a) experimentally altering how the fundraiser contacted potential donors in $2014$ and (b) analyzing individuals' behavior in $2014$ and $2015$.

\addvspace{0.1cm}\noindent\textbf{Control group.}\hspace{0.2cm}
Individuals in our experiment's control group received the standard solicitation letter, the contents of which remained unchanged from the pre-experiment years. Particularly, the letter highlighted the fundraiser's cause and asked recipients for a donation. To lower transaction costs, the fundraiser distributed the solicitation letter together with a remittance slip, prefilled with the fundraiser's bank account and the donor's name. In the pre-experiment years, potential donors received identical transaction forms.%
\footnote{For years, donations in the context of the fund drive could be made exclusively via bank transfer.}

\begin{figure}[tbp!]
	\begin{center}
		\caption{The gift consisting of the three folded cards and envelopes}
		\label{fig:sample_gift}
		\bigskip
		\includegraphics[scale=0.08]{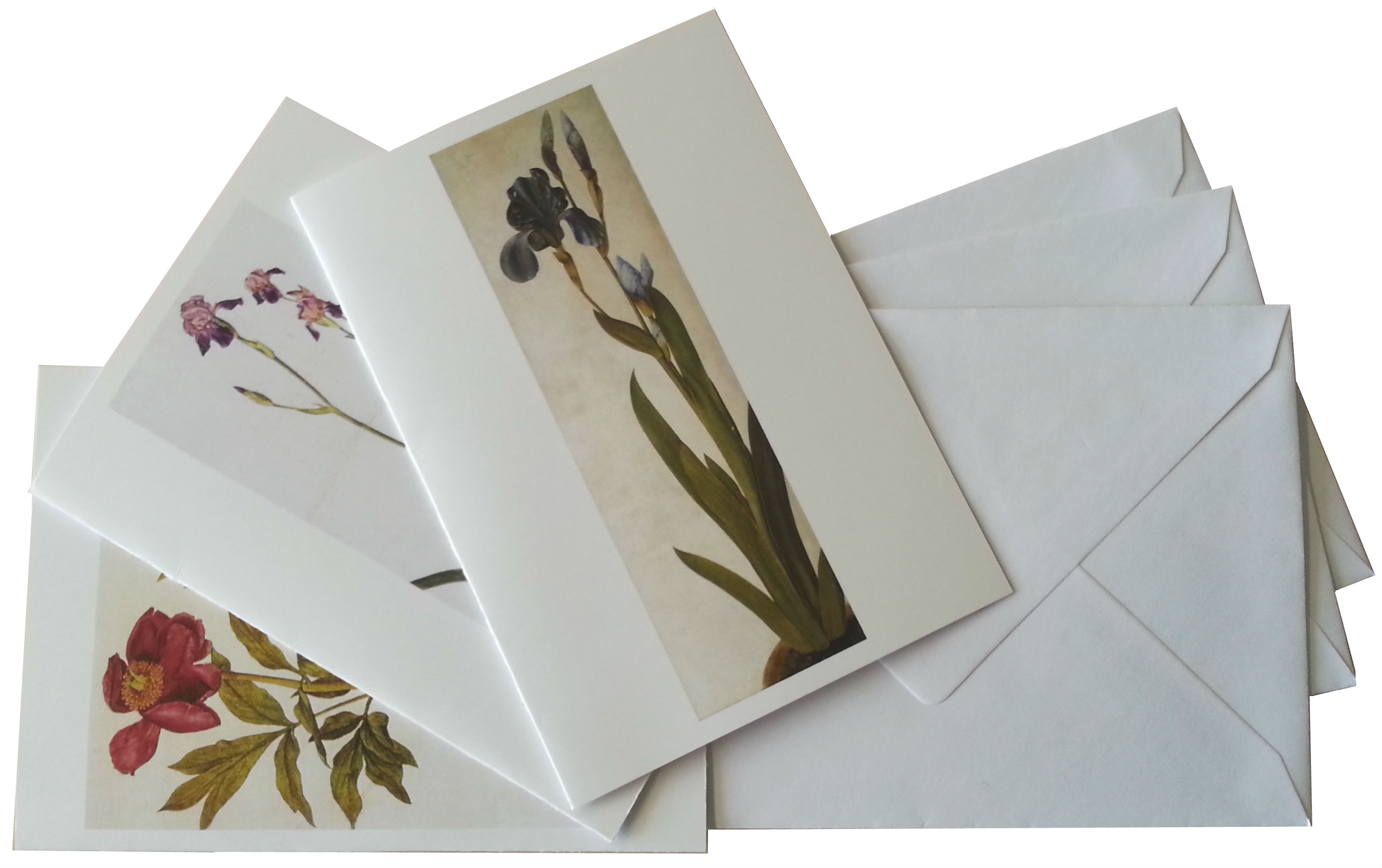}
		
		\vspace{0.2cm}
		\parbox{\textwidth}{\footnotesize{
				\textbf{Notes:} The gifts consisted of three different folded cards showing flower motifs from paintings of Albrecht D\"{u}rer plus three envelopes.}}
	\end{center}
\end{figure}

\addvspace{0.1cm}\noindent\textbf{Gift treatment.}\hspace{0.2cm}
Our design of the gift treatment closely follows \cite{Fal2007}. In $2014$, individuals in this treatment received the solicitation letter together with an unconditional gift. The gift consisted of three envelopes paired with different folded cards, picturing Albrecht D\"{u}rer's ``immaculate flower studies'' (see Figure \ref{fig:sample_gift}). Further, we added one sentence to the solicitation letter, stating that the fundraiser ``would like to provide the included folded cards as a gift.'' The total per-unit cost for mailing the control-group solicitation letter amounted to 0.43 euro (printing plus postage). In the gift treatment, the per-unit cost increased by $1.16$ euro (postcards and envelopes: $0.47$ euro; boxing and additional postage: $0.69$ euro). From $2015$ onward, all individuals in the sample received a solicitation letter very similar to the one distributed in the pre-experiment years.

\addvspace{0.1cm}\noindent\textbf{Sample.}\hspace{0.1cm}
In $2014$, $26\%$ of the urban area's population were members of the Catholic Church. We drew a sample from this population consisting of $2{,}354$ warm-list individuals (individuals who had donated at least once before the experiment) and $17{,}425$ cold-list individuals (individuals who had never donated before). These individuals were then randomly allocated to the control and treatment groups, exploiting a stratified randomization scheme. Particularly, we assigned $1{,}180$ of the warm-list individuals to the gift treatment and $1{,}174$ to the control group.%
\footnote{The strata were defined based on list (warm vs.~cold), gender, household type indicators, quintiles of individuals' predicted baseline willingness to give, and quintiles of age. To construct a proxy for the baseline willingness to give in the treatment year, we first regressed an indicator variable for giving in the year before the experiment on indicator variables for further lags of the giving indicator. We then used the estimated model to predict the probability of giving in the treatment year (out of sample).}
By contrast, $2{,}283$ cold-list individuals received the treatment, while $15{,}142$ were part of the control group.

\addvspace{0.1cm}\noindent\textbf{The setting's benefits.}\hspace{0.2cm}
Our setting serves as a suitable testing ground for machine-learning-based optimal targeting. First, as the fundraiser did not employ any targeting strategies before the experiment, the setting offers a clean environment to study our machine-learning approach's potential. Second, it provides rich data that not only allow us to estimate powerful targeting rules, but also enable us to test which type of data are especially beneficial for machine-learning-based optimal targeting (see the following description of the data). Third, because the fundraiser contacts all church members exhaustively, the setting offers the possibility to study the cold and warm lists separately. We, hence, can not only examine the optimal targeting of gifts among past donors, but also explore whom to target in the process of acquiring new donors. Fourth, because we were able to gather data for two postexperiment years, the setting allows us to study if the estimated optimal targeting rule increases total donations or pulls forward donations from $2015$ to $2014$. Also, we can examine the rule's external validity. Fifth, because religious giving dominates the charitable giving landscape \citep{List2011}, targeting is particularly relevant in this context.%
\footnote{For example, in Germany, church-related causes benefit the most from private giving: They receive approximately $35\%$ of total private giving \citep{Sepnd2016}. No other type of cause benefits from a similarly high share of total donations. The numbers for the United States are very similar \citep{And2013}.}

\subsection{The Data}
\noindent\textbf{Data sources.}\hspace{0.2cm}
Our study draws on two separate, comprehensive data sets. The first set of data includes administrative records provided by the Catholic Church. The records hold a number of socioeconomic characteristics, such as gender, marital status, and age. Furthermore, they contain individual-specific information on donations for the years $2006$--$2015$. Accordingly, we observe all potential donors' donation histories for eight pre-experiment years and their donations in the first two years after the experiment. Our second data source is Google Maps. Specifically, we used the Google Maps API to collect geospatial information on economic and cultural facilities near each individual's residence. We then merged this data with the administrative records based on postal addresses. In particular, we collected the number of restaurants, supermarkets, medical facilities, cultural facilities, and churches within $300$ meters of each the home address. We also web-scraped the distance from the home address to the central train station, city hall, main church, and airport. Furthermore, we retrieved the elevation of the home address. The main reason for using these geospatial characteristics is that they are readily available to charities and are powerful proxies for income and other socioeconomic characteristics \citep{don19, glae18,glae20} that likely explain response heterogeneity. Taken together, our algorithms for optimal targeting rely on three types of input data: (a) socioeconomic information, (b) information on past donation behavior, and (c) publicly available geospatial information. Note that charities that manage to collect even more comprehensive datasets could further improve machine-learning-based optimal targeting.

\addvspace{0.1cm}\noindent\textbf{Descriptive statistics.}\hspace{0.2cm}
Table \ref{desc_out} in Online Appendix \ref{app:descriptives} reports the descriptive statistics for the donation amount and the donation probability. Furthermore, supplementary tables either study the balance of observable characteristics across the warm and cold lists (see Table \ref{desc} in Online Appendix \ref{app:descriptives}) or the control and treatment groups (see Tables \ref{bal_warm} and \ref{bal_cold}). Several features of the data stand out. First, unsurprisingly, cold-list individuals donated much less in the first year after the experiment (average donation: $0.18$ euro) compared to warm-list individuals (average: $16.02$ euro). Their donation probability is also much lower. Similar results emerge when summing up the donations made in the first two years after the experiment. Second, donations are highly right-skewed and have excess kurtosis, highlighting that few donors give very large gifts. The following analysis highlights that, despite this data feature complicating our prediction task, we are nevertheless able to estimate effective optimal targeting rules. Third, cold list and warm list individuals differ in socioeconomic characteristics (see Table \ref{desc}).%
\footnote{On average, cold list individuals are younger, have a higher likelihood of being single, and tend to have a shorter residency duration in the urban area. Individuals in the warm list donated an average of four times, with a total donation amount of $126$ euro over the eight years before the experiment. By construction, cold list individuals have a donation history of zero donations. Individuals in the cold list live, on average, closer to the city center (closer to city hall and the central station) than individuals on the warm list. Close to the home address (within $300$ meters), cold-list individuals have, on average, more access to restaurants, supermarkets, medical and cultural facilities, and churches than warm-list individuals.}
This observation suggests that individual characteristics might explain heterogeneous giving behavior. Fourth, the observable characteristics are balanced across the treatment and control groups (see Tables \ref{bal_warm} and \ref{bal_cold}).

\section{Empirical Strategy}\label{sec:emprical_strategy}
This section describes the estimation and identification strategy of machine-lear\-ning-based optimal targeting rules and how to assess the rules' out-of-sample performance. We start by introducing conditional average treatment effects (called CATEs) in Subsection \ref{subsec_cate}. The CATEs formally describe heterogeneous treatment effects as a function of observable characteristics. In Subsection \ref{subsec_targeting_rules}, we then introduce binary optimal targeting rules that are related to the continuous CATEs. These rules maximize the charity's net donations by assigning individuals to either the targeted group or the untargeted group. Recall that our approach directly estimates the optimal targeting rule. This feature distinguishes it from approaches that first estimate the continuous CATEs and then derive the rules as nonlinear transformations of these effects. In the final step, we discuss our classification approach in Subsection \ref{subsec_estimation}. In Subsection \ref{subsec_estimation}, we further discuss how we measure our targeting rules' out-of-sample performance compared to benchmark rules.

\subsection{Conditional Average Treatment Effects}\label{subsec_cate}
\noindent\textbf{Notation.}\hspace{0.2cm}
We use the potential-outcome framework \citep{rub74} to describe the parameters of interest. The potential-outcome framework is useful when studying targeting because it allows us to describe an individual's reaction under different (counterfactual) treatment conditions. The treatment variable $D_i$ indicates whether a fundraising gift was sent to individual $i$ (for $i = 1, \ldots, N$), with
\begin{equation*}
D_i = \left\{ \begin{array}{rl} 1 & \mbox{ when a gift was sent, and} \\ -1 & \mbox{ otherwise.} \end{array} \right.
\end{equation*}
$Y_{i}(1)$ denotes the potential donations in response to the solicitation letter with a fundraising gift. $Y_{i}(-1)$ denotes the potential donations in response to the letter without the gift.

\addvspace{0.1cm}\noindent\textbf{Causal effects.}\hspace{0.2cm}
Using the previous notation, the individual causal effects are
\begin{equation*}
\delta_{i} = Y_{i}(1)-Y_{i}(-1).
\end{equation*}

In an ideal world, the charity would know $\delta_{i}$. It could then (exclusively) assign the gift to individuals for whom $\delta_{i}$ exceeds the gift's cost. However, because $Y_{i}(1)$ and $Y_{i}(-1)$ cannot be observed simultaneously, the fundamental problem of causal analysis is that $\delta_{i}$ is unobservable. Nevertheless, it is possible to identify and estimate group averages of $\delta_{i}$. For example, the average treatment effect (ATE), $\delta = E[\delta_i]= E[Y_{i}(1)-Y_{i}(-1)]$, is the expected average effect of the gift on donations. Moreover, there might be effect heterogeneity with regard to observable characteristics, $X_{i}$, which allows researchers to identify even finer-grained subgroup-specific effects. For example, \cite{andr01} and \cite{andr03} show that men and women differ in their donation behavior. In this vein, the CATE describes the association between an individual's characteristics $x$ and the expected effect of sending the fundraising gift to the individual:
\begin{equation*}
\delta(x) = E[\delta_{i}|X_{i}=x] = E[Y_{i}(1)-Y_{i}(-1)|X_{i}=x].
\end{equation*}

\addvspace{0.1cm}\noindent\textbf{Identification of causal effects.}\hspace{0.2cm}
The ATEs and CATEs are identified under the stratified experimental design and the stable unit treatment value assumption (SUTVA):
\begin{equation*}
Y_{i} = Y_i(-1) + \frac{1+D_i}{2} \left(Y_i(1)-Y_i(-1) \right),
\end{equation*}
(see proof in Online Appendix \ref{app:CATEs}). The proof distinguishes between strata characteristics, which we call $Z_i$, and heterogeneity variables, $X_i$. Due to stratified randomization, we need to account for the strata characteristics to achieve identification. In contrast, we do not require the heterogeneity variables for identification. They, however, are potentially associated with heterogeneous effects of the gift.

\subsection{Targeting Rules}\label{subsec_targeting_rules}
\noindent\textbf{Optimal targeting rule.}\hspace{0.2cm}
A targeting rule, $\pi(X_i) \in \{-1, 1\}$, is a deterministic function that assigns the gift to prospective donors based on their observable characteristics, $X_i$. Under the rule, individuals with $\pi(X_i)=1$ receive the solicitation letter with the gift, and individuals with $\pi(X_i)=-1$ receive the solicitation letter without the gift. The purpose of the optimal targeting rule is to maximize the expected net donation $P(\cdot)$ of the fundraising campaign, defined as the expected donation minus the gift's variable costs. Formally, the expected net donation is
\begin{equation} \label{obj1}
P(\pi(X_i)): = E\left[Y_i(\pi(X_i))- \frac{1+\pi(X_i)}{2}c \right] ,
\end{equation}
where $Y_i(\pi(X_i))$ is the donation amount of individual $i$ under the rule $\pi(X_i)$ and $c$ are the variable costs of the gift. We ignore fixed costs, as they do not alter the targeting rule.

\addvspace{0.1cm}\noindent\textbf{Benchmarks rules.}\hspace{0.2cm}
To evaluate the gains of optimal targeting, we compare the expected net donations under the optimized rule $P(\pi(X_i))$ to the expected net donations under three benchmarks: a rule that assigns the gift to everybody (all-gift benchmark), a rule that assigns the gift to no one (no-gift benchmark), and a rule with random allocation (random-gift benchmark). First, we can contrast the optimal targeting rule to the all-gift benchmark $\pi(X_i)=\pi_1=1$. For this benchmark, the expected net donation is $P(\pi_1)= E[Y_i(1)]- c$. Consequently, the excess net donation of the optimal targeting rule compared to this benchmark is
\begin{equation*}
Q_1(\pi(X_i)):= P(\pi(X_i)) - P(\pi_{1})= E\left[ \frac{\pi(X_i)-1}{2} \left( \delta_i - c \right) \right].
\end{equation*}

Second, equivalently, we compare the optimal rule to the no-gift benchmark $\pi(X_i)=\pi_{-1}=-1$, under which the expected donations are $P(\pi_{-1})= E[Y_i(-1)]$. Thus, relative to this benchmark, optimal targeting increases net donations by
\begin{equation*}
Q_{-1}(\pi(X_i)):= P(\pi(X_i)) - P(\pi_{-1})=
E \left[ \frac{1+\pi(X_i)}{2} \left( \delta_i - c \right) \right].
\end{equation*}

Third, we consider the random-gift benchmark, $\pi_R$, under which each individual has a $50\%$ probability of receiving the gift. Given that $\pi_R$ triggers the expected net donation of $P(\pi_R)= 1/2 \cdot (E[Y_i(1) + Y_i(-1)]- c)$, the excess net donation of the optimal rule is
\begin{equation} \label{ra}
\begin{array}{rl}
Q_R(\pi(X_i)):=& \displaystyle P(\pi(X_i)) - P(\pi_{R})= \frac{1}{2} E\left[  \pi(X_i) \left( \delta_i - c \right) \right]\\ =& \displaystyle [Q_1(\pi(X_i)) + Q_{-1}(\pi(X_i))]/2. \end{array}
\end{equation}
The random rule can be viewed as a default option when no information about the effectiveness of the fundraising instrument is available, and the fundraiser has no preferences about the allocation of the instrument.

Two further points are of note. First, the optimal targeting rule which maximizes the net donations $P(\cdot)$, also maximizes $Q_{1}(\cdot)$, $Q_{-1}(\cdot)$, and $Q_{R}(\cdot)$. The reason is that $P(\pi_{1})$, $P(\pi_{-1})$, and $P(\pi_{R})$ are constant. Second, for the estimation of the optimal targeting rule, we maximize the sample analog of \eqref{ra}. Because we do not observe the individual causal effects, $\delta_i$, which we need to determine \eqref{ra}, we first discuss how to approximate these parameters.

\subsection{Estimation}
\label{subsec_estimation}
To introduce our estimation strategy of the optimal targeting rule, we proceed in two steps. In the first step, we discuss augmented inverse probability weighting (AIPW) to estimate an approximation of the individual causal effects. In the second step, we show how to use the AIPW score to estimate the optimal targeting rule.

\addvspace{0.1cm}\noindent\textbf{Augmented inverse probability weighting.}\hspace{0.2cm}
An essential ingredient for the optimal targeting rule is $\delta_i$. As we mentioned before, $\delta_i$ is unobservable and cannot be estimated directly. However, an approximation score of $\delta_i$ can be sufficient to estimate the optimal targeting rule. The AIPW score,
\begin{equation*}
\Gamma_i =  {\mu}_1(Z_{i}) - {\mu}_{-1}(Z_{i}) + \frac{1+D_{i}}{2} \cdot \frac{Y_{i} - {\mu}_{1}(Z_{i}) }{{p}(Z_{i})} + \frac{D_{i}-1}{2} \cdot\frac{Y_{i} - {\mu}_{-1}(Z_{i}) }{1-{p}(Z_{i})},
\end{equation*}
which depends on the experimental strata characteristics $Z_i$, is an example of such an approximation score.%
\footnote{Alternatively, \cite{kit18} suggest inverse probability weighting scores, and \cite{bey09} propose offset weighting scores.} 
The so-called nuisance parameters are the conditional expectations of the donations, ${\mu}_{1}(z) = {E}[Y_i |D_i=1,Z_i=z]$ and ${\mu}_{-1}(z) = {E}[Y_i |D_i=-1,Z_i=z]$, and the conditional probability that the gift was sent ${p}(z) = {Pr}(D_i= 1|Z_i=z)$. The latter is often called the propensity score. Under the SUTVA and the experimental design, the expected value of the AIPW score identifies the ATE $\delta = E[\Gamma_i]$.%
\footnote{Note that the nuisance parameters, ${\mu}_{1}(z) = {E}[Y_i |D_i=1,Z_i=z]= {E}[Y_i(1) |Z_i=z]$ and ${\mu}_{-1}(z) = {E}[Y_i |D_i=-1,Z_i=z]= {E}[Y_i(-1) |Z_i=z]$, equal conditional expectations of the potential donations with and without the gift under the SUTVA and the experimental design.}
The conditional expectations of the AIPW score identify the CATEs, $\delta(x)= E[{\Gamma}_i|X_i=x]$. For completeness, we sketch the identification proofs for the AIPW score in Online Appendix \ref{app:AIPW}.

\addvspace{0.1cm}\noindent\textbf{Estimating the AIPW score.}\hspace{0.2cm}
We can estimate the AIPW score as follows. First, we estimate the nuisance parameters. In this step, we obtain the estimated conditional expectations of the potential donations with and without the gift by $\hat{\mu}_{1}(z)$ and $\hat{\mu}_{-1}(z)$ and the estimated propensity score by $\hat{p}(z)$. Second, we plug the estimated nuisance parameters into the estimator of the AIPW score, $\hat{\Gamma}_i=  \hat{\Gamma}_i(1)-\hat{\Gamma}_i(-1)$, with
\begin{equation*}
\hat{\Gamma}_i(1)=  \hat{\mu}_1(Z_{i})  + \frac{1+D_{i}}{2} \cdot \frac{Y_{i} - \hat{\mu}_{1}(Z_{i}) }{\hat{p}(Z_{i})}  ,
\end{equation*}
and
\begin{equation*}
\hat{\Gamma}_i(-1)=   \hat{\mu}_{-1}(Z_{i})  - \frac{D_{i}-1}{2} \cdot\frac{Y_{i} - \hat{\mu}_{-1}(Z_{i}) }{1-\hat{p}(Z_{i})}.
\end{equation*}

The corresponding average-treatment-effect estimator
\vspace{-0.3cm}\begin{equation} \label{aipw}
\hat{\delta}= \frac{1}{N} \sum_{i=1}^{N} \hat{\Gamma}_i
\end{equation}
is consistent, asymptotically normal, and semiparametrically efficient under the requirement that the nuisance parameter estimators are consistent and converge sufficiently fast \citep[e.g.,][]{Chernozhukov2017,Robins1994}. In our application, we have precise information about the stratification process. Therefore, we use parametric nuisance parameter estimators which satisfy the requirements. In particular, we use a logit to estimate the propensity score and OLS to estimate the conditional expectations of the potential donations with and without the gift (Table \ref{nuisance} in Online Appendix \ref{app:nuisance} reports the estimated coefficients of the different models).%
\footnote{Online Appendix \ref{app:sensitivity} presents two robustness tests regarding the specification of the nuisance parameters. First, it replaces the estimated propensity score with the population propensity score. Second, it estimates all nuisance parameters with cross-fitted logit-lasso models instead of conventional estimators. The results are quantitatively identical to our baseline results (see Tables \ref{pop_weight} and \ref{lasso_nuisance} in Online Appendix \ref{app:sensitivity}).}


Note that we could alternatively estimate the ATEs with a OLS regression model. However, in contrast to OLS, the AIPW estimator does not impose any linearity assumptions and does not restrict effect heterogeneity. The latter is particularly relevant for the targeting approach. Having said this, we show in Section \ref{sec:results} that the OLS and AIPW estimates of the ATEs are similar.

\addvspace{0.1cm}\noindent\textbf{Estimating the optimal targeting rule.}\hspace{0.2cm}
For the estimation of the optimal targeting rule, \cite{ath21} propose replacing the unobservable individual causal effect in the sample analog of \eqref{ra}, $\delta_i$, with the estimated AIPW score, $\hat{\Gamma}_i$:
\begin{equation} \label{ob2}
\pi^* =  \mbox{argmax}_{\pi} \left\{  \frac{1}{2N} \sum_{i=1}^{N} \pi(X_i)(\hat{\Gamma}_i - c ) \right\}.
\end{equation}
Alternatively, the objective function \eqref{ob2} can be formulated as the weighted classification estimator
\begin{equation} \label{obj3}
\pi^* =  \mbox{argmax}_{\pi} \left\{  \frac{1}{2N} \sum_{i=1}^{N} \pi(X_i) \cdot \mbox{sign}(\hat{\Gamma}_i-c) \cdot |\hat{\Gamma}_i - c | \right\},
\end{equation}
where $(\hat{\Gamma}_i-c)=\mbox{sign}(\hat{\Gamma}_i-c) \cdot |\hat{\Gamma}_i - c |$ \citep[see, e.g.,][]{bey09,zadr03,zhao12}. This estimator aims to classify the sign of the net donation effects and weigh each observation by $|\hat{\Gamma}_i - c |$. The objective function is maximized when the signs of $\pi(X_i)$ and $(\hat{\Gamma}_i-c)$ are equal. If some signs differ, misclassifications of individuals who respond strongly (i.e., individuals with large weights) reduce the net donations more than misclassification of individuals who do not respond strongly to the gift (i.e., individuals with small weights). Accordingly, the optimal targeting estimator should prioritize individuals with large weights. Because the estimated optimal targeting rule is a deterministic function of the observable characteristics $X_i$, there is an implicit connection between optimal targeting and the CATEs, even though we estimate the optimal targeting rule directly (without estimating the CATEs first). %

The main result of \cite{ath21} that enables estimation of \eqref{obj3} is that, when the complexity of the estimator is restricted, the optimal targeting rule $\pi^*$ achieves asymptotically minimax-optimal regret \citep[][]{man04}. Along these lines, in principle, any restricted weighted classification estimator could be used to estimate \eqref{obj3}. We, however, follow \cite{ath21} and use shallow decision trees to estimate the optimal rule. Trees partition the sample into mutually exclusive strata based on the heterogeneity characteristics, $X_i$. Furthermore, the tree depth restricts the complexity of the estimated optimal targeting rule, which makes trees suitable estimators in our context. In our main specifications, we follow \cite{zho2019} and use exact policy-learning trees, with a search depth of two, to estimate the optimal targeting rule.%
\footnote{For implementation, we use the \texttt{R} package \texttt{policytree} \citep{sver20}.}

\addvspace{0.1cm}\noindent\textbf{Advantages of decision trees.}\hspace{0.2cm}
It is  possible to use standard estimators, such as a weighted logit regression, to estimate \eqref{obj3}. However, decision trees have several advantages compared to logit regressions for the estimation of optimal targeting rules. They select the relevant heterogeneity characteristics in a data-driven way by balancing the bias-variance trade-off. This feature is particularly useful when we have no \textit{a priori} domain knowledge about the relevant characteristics. Even if we know the relevant characteristics, there might be several highly correlated measures of these characteristics, and it may be \textit{a priori} unclear which are the most relevant. For example, in our application, several geospatial characteristics are highly correlated, and there is little \textit{a priori} guidance on which should be used. In the extreme case, including too many highly correlated characteristics in a logit regression could cause multicollinearity problems. Furthermore, it is typically unclear how flexible the empirical model should be with regard to nonlinear and interaction terms. Trees can automatically incorporate nonlinear and interaction terms of the different characteristics without precoding. This feature minimizes the risk of overlooking important heterogeneities. Having said this, we study the sensitivity of our results to different estimation methods for the targeting rule in Section \ref{subsec:robustness}. In particular, we consider logit, logit lasso, classification and regression trees (CART), and classification-forest estimators.

\addvspace{0.1cm}\noindent\textbf{Estimating the gains of targeting.}\hspace{0.2cm}
Once we have estimated the optimal targeting rule, $\pi^*$, we can apply the sample analogy principle to estimate the gains of targeting relative to the benchmarks: \vspace{-.4cm}
\begin{align*}
\hat{P}(\pi^*(X_i))&=  \frac{1}{N} \sum_{i=1}^{N} \left(   \hat{\Gamma}_i(\pi^*(X_i)) - \frac{1+\pi^*(X_i)}{2} c \right) ,\\
\hat{Q}_1(\pi^*(X_i))&=  \frac{1}{N} \sum_{i=1}^{N}  \frac{\pi^*(X_i)-1}{2} \left( \hat{\Gamma}_i - c \right),\\
\hat{Q}_{-1}(\pi^*(X_i))&=
\frac{1}{N} \sum_{i=1}^{N}  \frac{1+\pi^*(X_i)}{2} \left( \hat{\Gamma}_i - c \right) \mbox{, and}\\
\hat{Q}_{R}(\pi^*(X_i))&=
\frac{1}{2N} \sum_{i=1}^{N}  \pi^*(X_i) \left( \hat{\Gamma}_i - c \right) .
\end{align*}
These estimators are consistent, asymptotically normal, and semiparametrically efficient \citep[see][]{unify18}.

We estimate the gains of targeting using a cross-validation procedure. Our procedure randomly partitions our data into $K=20$ equally sized samples. We then use $K-1$ partitions to estimate the targeting rule $\pi^*$ and calculate $\hat{P}(\pi^*(X_i))$, $\hat{Q}_1(\pi^*(X_i))$, $\hat{Q}_{-1}(\pi^*(X_i))$, and $\hat{Q}_R(\pi^*(X_i))$ in the retained partition. We repeat this procedure, discarding each of the $K$ partitions once. In this way, we use the entire dataset efficiently. Finally, we report the average values of $\hat{P}(\pi^*(X_i))$, $\hat{Q}_1(\pi^*(X_i))$, $\hat{Q}_{-1}(\pi^*(X_i))$, and $\hat{Q}_R(\pi^*(X_i))$ over all $20$ partitions. The cross-validation approach allows us to assess the estimated targeting rules' out-of-sample performance. It also addresses the concern that the targeting rule reflects spurious relationships and overstates the success of targeting due to overfitting.

\section{Results}
\label{sec:results}
This section presents our results. Subsection \ref{subsec:ATEs} discusses the ATEs of the gift on donations, and Subsection \ref{subsec:heterogeneity} explores the heterogeneity of the effects. The section proceeds by discussing the effectiveness of our optimal targeting approach in Subsection \ref{subsec:targeting} before describing several properties of the estimated optimal targeting rule in Subsection \ref{subsec:characteristics}. Subsection \ref{subsec:datasources} outlines which characteristics are sufficient to increase profits significantly through machine-learning-based optimal targeting. Finally, Subsection \ref{subsec:stability} examines the stability of the estimated optimal targeting rule, and Subsection \ref{subsec:robustness} explores the robustness of our results to alternative estimators.

\subsection{Average Effects of Gifts on Donations}
\label{subsec:ATEs}
To facilitate comparison to the literature studying the effects of unconditional gifts on donations, our first step is to estimate the ATE of the gift on donations. Table \ref{ate} reports three different estimates, focusing on behavior in the first year after the experiment: estimates from unconditional OLS regressions (Columns 1 and 4), estimates from conditional OLS regressions (Columns 2 and 5), and  estimates from the previously introduced AIPW estimator (Columns 3 and 6).%
\footnote{In contrast to the OLS regressions, the AIPW estimator relies on fewer functional-form assumptions, allows for heterogeneous treatment effects, and is more robust to misspecification.}
Columns 1--3 cover the warm list and Columns 4--6 the cold list.

\addvspace{0.1cm}\noindent\textbf{Average treatment effects for the warm list.}\hspace{0.2cm}
Two observations characterize the responses in the warm list. First, the gift increased average donations by $1.21$--$1.24$ euro, though the effects are not statistically significant  (see Row A in Table \ref{ate}).%
\footnote{This result is in line with \cite{lan10}, who also report insignificant effects for the warm list.}
Notably, these estimates do not account for the cost of the gift (which were $1.16$ euro). Second, when accounting for the costs, we find small and insignificant net-of-cost effects between $0.05$ euro and $0.08$ euro, depending on the chosen estimator (see Row B in Table \ref{ate}). The small values imply that we neither find evidence for the hypothesis that the gift treatment was, on average, profitable (i.e., increased average donations by more than the costs) nor that it resulted in a net loss for the fundraiser. The two observations are insightful from a targeting perspective. To see why, note that we can think of the treatment effects as driven by a change from the benchmark targeting rule where no one receives the gift (control group) to the one where everybody receives the gift (treatment group).%
\footnote{More formally, the expected difference in net donations between the benchmark rules $\pi_{1}$ and $\pi_{-1}$ is $P(\pi_{1})-P(\pi_{-1})=E[\delta_i]-c$.}
Along these lines, the insignificant net-of-cost effects speak against the hypothesis that the all-gift benchmark outperforms the no-gift benchmark in terms of available funds.

\begin{table}[tbp!]
	\footnotesize \singlespacing
	\begin{center}
		\caption{Average treatment effects of the gift on donations} \label{ate}
		\begin{tabularx}{\textwidth}{Xccccccc} \toprule
			& \multicolumn{3}{c}{Warm list} && \multicolumn{3}{c}{Cold list} \\
			& OLS & OLS & AIPW &&  OLS & OLS & AIPW\\
			\cline{2-4} \cline{6-8}  & (1)   & (2)  & (3) & & (4)  & (5)  & (6)    \\\midrule
			A. Average treatment effects   & 1.24 &  1.21 &  1.22 &  & 0.19*** &  0.19*** & 0.19* \\
			& (1.25) & (1.16) & (1.15)& & (0.07) & (0.07) & (0.10) \\\\
			B. Average treatment effects  & 0.08 &  0.05 &  0.06 &  & -0.97*** &  -0.97*** & -0.97*** \\
			net of costs       & (1.25) & (1.16) & (1.15)& & (0.07) & (0.07) & (0.10) \\ \midrule
			Strata controls & No   & Yes & Yes & & No   & Yes & Yes \\ \bottomrule
		\end{tabularx}
	\end{center}
	\parbox{\textwidth}{\footnotesize{\textbf{Notes:} This table shows the estimated ATEs of the gift treatment on donations. The first set of estimates uses the amount donated in the first year after the gift as an outcome variable (euro). The second set of estimates additionally subtracts the gift's cost from the donation amount. We report results for the following specifications: unconditional OLS (Columns 1 and 4), OLS with strata control variables (Columns 2 and 5), and AIPW (Columns 3 and 6). Because the AIPW model allows for heterogeneous treatment effects, this model represents our preferred specification. Standard errors are in parenthesis. ***/**/* indicate statistical significance at the 1\%/5\%/10\% level.}}
\end{table}

\addvspace{0.1cm}\noindent\textbf{Average treatment effects for the cold list.}\hspace{0.2cm}
Very different results emerge for the cold list. The ATEs are significant (note the larger sample size), but much smaller, and amount to just $0.19$ euro (Row A). As a result, when accounting for costs, the all-gift benchmark rule would result in a significant loss of $0.97$ euro per donor, compared to the no-gift benchmark rule (Row B). Accordingly, considering only these two benchmark rules, we find that the more profitable strategy is to send the gift to no one in the cold list.%
\footnote{The previous literature has reported similar results in the past. For example, \cite{ALP2008} show that gifts increase donations, but the increase is insufficient to cover the gift's costs.}

Taken together, we conclude that naive targeting of gifts to all individuals in the warm and cold list does not increase our fundraiser's net donations. Such a strategy would even likely result in a net loss as there are many more cold-list than warm-list individuals. Based on these insights, one might be tempted to conclude that our fundraiser can never increase net donations with the gift. In the following, we, instead, show that a more targeted gift campaign can be very successful.

\subsection{Heterogeneous Treatment Effects}
\label{subsec:heterogeneity}
This subsection provides a descriptive analysis suggesting that the charity can likely increase raised net donations by deviating from the no-gift and all-gift benchmarks. For that purpose, we demonstrate that the effects of the gift are heterogeneous. To motivate the analysis, note that in the absence of heterogeneous treatment effects, the charity cannot benefit from targeting rules that are more flexible than the benchmarks. Intuitively, as all individuals respond similarly to the gift, the optimal targeting rule would either correspond to the no-gift or all-gift benchmark. However, with effect heterogeneity, some individuals may increase their donations by more and others by less than the gift's cost. If this is the case, the charity's optimal strategy would be to deviate from the benchmarks and target a subset of individuals only.

\addvspace{0.1cm}\noindent\textbf{Sorted-effects approach.}\hspace{0.2cm}
As a first step, we use the CATE-based sorted-effects approach of \cite{sort} to study descriptively if there is sufficient heterogeneity for optimal targeting. This method allows us to visualize the distribution of the effects of the gift on donations while reporting confidence intervals that account for multiple testing. Particularly, we specify a linear OLS regression including all observable characteristics plus interactions with the treatment dummy. Using this model, we estimate the CATE for each individual and report the percentiles of the estimated CATEs (labeled \textit{sorted effects}).%
\footnote{Statistical inference is based on a multiplier bootstrap \citep[see][for details]{sort}.}
We then examine if the sorted-effects model shows heterogeneous effects below and above the gift's costs to investigate if deviations from the two benchmarks are likely beneficial. Note, however, that we only use the sorted-effects approach to assess the potential of optimal targeting descriptively (i.e., we do not use it to identify targets). Instead, we estimate the optimal targets using a machine-learning approach explicitly developed for this purpose.

\begin{figure}[tbp!]
	\begin{center}
		\caption{Sorted effects} \label{spe}
		\begin{subfigure}{0.49\textwidth}
			\includegraphics[width=\textwidth]{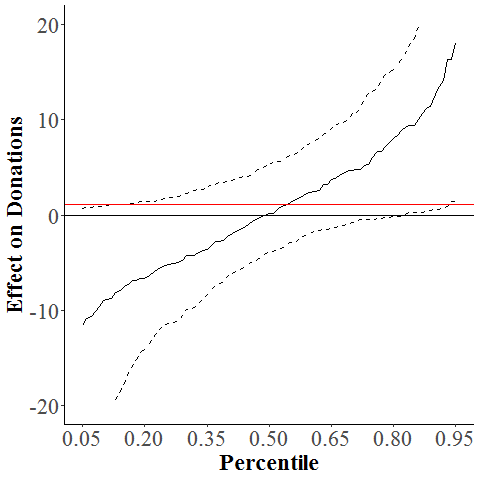}
			\caption{Warm list}
		\end{subfigure}
		\begin{subfigure}{0.49\textwidth}
			\includegraphics[width=\textwidth]{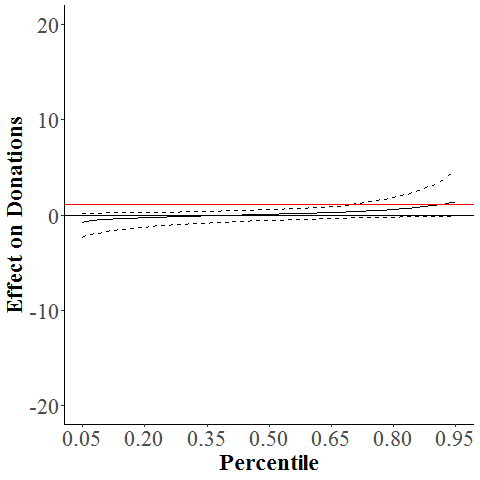}
			\caption{Cold list}
		\end{subfigure}
		\parbox{\textwidth}{\footnotesize{\textbf{Notes:} This figure shows the heterogeneity of the effect of the gift on the donation amount. To that end, it sorts the estimated conditional average treatment effects by size and plots the size of the treatment effect in euro (vertical axis) against the percentiles of the effect size (horizontal axis). The red horizontal line represents the cost of the gift (1.16 euro). The solid line depicts the sorted effects. We report results between the $5$ and $95$ percentiles. The dashed lines report uniformly valid $95\%$ multiplier-bootstrap confidence intervals ($500$ replications).}}
	\end{center}
\end{figure}

\addvspace{0.1cm}\noindent\textbf{Heterogeneity in the warm list.}\hspace{0.2cm}
Panel (a) in Figure \ref{spe} depicts the heterogeneity of the treatment effect on the donation amount for the warm list (solid line). It sorts the estimated CATEs by size and plots the size of the treatment effect in euro (vertical axis) against the percentiles of the effect size (horizontal axis). The red horizontal line represents the cost of the gift (1.16 euro).

The figure reveals substantial treatment-effect heterogeneity.%
\footnote{One might be interested in whether the size of the treatment effects correlates with observable characteristics. Tables \ref{high_low_10_warm} and \ref{high_low_10_cold} in Online Appendix \ref{app:sorted} report the mean values of all characteristics for the groups with the $10\%$ largest and the $10\%$ smallest sorted effects. In the warm list, the individuals with the $10\%$ largest effects tend to have donated less before the experiment and to live at a lower altitude than individuals with the $10\%$ smallest effects, although the effects are insignificant. In the cold list, the individuals with the 10\% largest effects tend to live significantly closer to the city center than individuals with the 10\% smallest effects.}
For some individuals, the treatment effects are positive, which is in line with the sequential-reciprocity hypothesis \citep{Duf2004,Fal2007}. For example, the $5\%$ most responsive individuals increase donations by more than $17.99$ euro in response to the gift. By contrast, at the fifth percentile, donations decrease by $11.59$ euro. Such an adverse effect points to the possibility that even a ``warm'' gift (folded cards) may change the donors' perception of the relationship with the fundraiser from a communal to an exchange norm \citep{yin2020coins}. The pronounced heterogeneity is interesting for at least two reasons. First, and most importantly, the heterogeneity indicates that machine-learning-based optimal targeting can be highly beneficial in the warm list: For $45\%$ of all individuals, the estimated effects exceed the cost of providing the gift. Thus, by targeting these individuals and not targeting donors with effects lower than the cost, the charity could substantially increase net donations. Second, the figure also reveals that the individual heterogeneity alone is powerful enough to account for the negative and positive effects reported in the literature.

\addvspace{0.1cm}\noindent\textbf{Heterogeneity in the cold list.}\hspace{0.2cm}
Panel (b) in Figure \ref{spe} highlights that the effect heterogeneity in the cold list is much smaller than in the warm list. For example, the donation amount decreases by $0.73$ euro at the fifth percentile and increases by $1.38$ euro at the $95$ percentile. Moreover, we find that the gift-induced increase in donations exceeds the costs only for individuals above the $92$ percentile in the cold list, and even for these individuals the size of the treatment effects is relatively small. 

\addvspace{0.1cm}\noindent\textbf{Heterogeneity test.}\hspace{0.2cm}
Rather than only presenting descriptive analyses, we also more formally test if there is detectable effect heterogeneity based on observed characteristics. For this purpose, we implement the best linear predictor method of \cite{Chernozhukov2020}. Particularly, following \cite{ath19b}, we use a test that relies on a causal forest estimator. Table \ref{het_test} in Online Appendix \ref{app:sorted} documents the findings.%
\footnote{Positive and statistically significant coefficients of the heterogeneity loadings provide evidence for detectable effect heterogeneity based on the observed characteristics. The causal forest approximates effect heterogeneity well when the coefficients of the heterogeneity loadings are close to one.}
In line with the previous results, the test detects significant effect heterogeneity in the warm list and no heterogeneity in the cold list. We conclude that the potential to increase net donations by subgroup-specific targeting is much lower in the cold list than in the warm list.

\begin{table}[tbph!]
	\footnotesize \singlespacing
	\begin{center}
		\caption{Out-of-sample performance of targeting rule in the warm list} \label{main_res_warm}
		\begin{tabularx}{\textwidth}{Xccccc} \toprule
			&    & Expected outcome value & \multicolumn{3}{c}{Optimal targeting vs. benchmarks}  \\
			&    & under optimal targeting  & all-gift & no-gift & random-gift \\
			\cline{3-6}   &     & (1)   & (2)   & (3)   & (4) \\ \midrule
			\multicolumn{6}{c}{\textbf{Panel A: Share of individuals that should receive the gift}} \\\midrule		
			A1. Share treated & 0.33 &  &  & &   \vspace{.15cm}\\	  \midrule		
			\multicolumn{6}{c}{\textbf{Panel B: Results for primary outcome variable}}\\\midrule
			\multicolumn{2}{l}{B1. Net donation amount}  & 17.61*** & 2.14*** & 2.20*** & 2.17*** \\
			(1st year)              &  & (0.97) & (0.82) & (0.81) & (0.58)  \vspace{.15cm}\\ \midrule
			\multicolumn{6}{c}{\textbf{Panel C: Results for secondary outcome variables}} \\\midrule
			\multicolumn{2}{l}{C1. Donation probability} & 0.503*** & 0.007 & 0.025** & 0.016* \\
			(1st year)     &       & (0.013) & (0.013) & (0.010) & (0.008) \\\\
			\multicolumn{2}{l}{C2. Net donation amount}  & 32.94*** & 2.33* & 3.75*** & 3.04*** \\
			(1st and 2nd year) &       & (1.66) & (1.41) & (1.41) & (1.00) \\\\
			\multicolumn{2}{l}{C3. Donation probability}  & 0.582*** & 0.001 & 0.017* & 0.009 \\
			(1st and 2nd year)     &       & (0.013) & (0.013) & (0.009) & (0.008) \\\bottomrule
		\end{tabularx}
	\end{center}
	\parbox{\textwidth}{\footnotesize{\textbf{Notes:} This table documents the out-of-sample performance of our estimated optimal targeting rule, focusing on the warm list. The goal of optimal targeting is to maximize donations, net of costs. Panel A reports the share of individuals that, according to the rule, should receive the gift. Panel B reports the expected consequences of our rule for net donations as our main outcome. Panel C, instead, focuses on secondary outcomes. Column 1 shows the expected value of the outcomes under optimal targeting. For example, we expect that, under optimal targeting, the donations, net of costs, would be $17.61$ euro. Columns 2--4 show how optimal targeting changes the outcomes relative to three benchmark scenarios: everybody receives the gift (Column 2), no one receives the gift (Column 3), and the gift is randomly assigned to half of the sample (Column 4). Donations are measured in euro. Standard errors are in parentheses. ***/**/* indicate statistical significance at the 1\%/5\%/10\% level.}}
\end{table}

\subsection{Effectiveness of Machine-Learning-Based Optimal Targeting}
\label{subsec:targeting}
This subsection evaluates if machine-learning-based optimal targeting allows us to exploit the documented response heterogeneity to increase net donations in the first year after the experiment. We first estimate optimal targeting rules \citep{ath21}. We then evaluate the effectiveness of the estimated rules in raising net donations by comparing their out-of-sample performance to our benchmarks. For that purpose, we use the cross-validation approach described in Section \ref{sec:emprical_strategy}.

\addvspace{0.1cm}\noindent\textbf{The estimated optimal targeting rule in the warm list.}\hspace{0.2cm}
Table \ref{main_res_warm} focuses on the warm list and documents how the estimated optimal targeting rule performs out of sample relative to the benchmarks. The table provides three sets of insights. First, Panel A reports that the estimated targeting rule recommends deviating from the no-gift and all-gift benchmarks. Specifically, it assigns the gift to $33\%$ of the warm-list individuals (Column 1).

Second, Panel B documents that, compared to our benchmarks, the charity would benefit substantially from applying the estimated optimal targeting rule. In the first year after the experiment, the average net donation under the estimated optimal targeting rule is $17.61$ euro (Column 1). This value implies that, under the estimated optimal targeting rule, the average donation, net of costs, is $2.14$ euro ($13.8\%$) higher than if everybody received the gift (Column 2), $2.20$ euro ($14.3\%$) higher than if no one received the gift (Column 3), and $2.17$ euro ($14.1\%$) higher than if the gift was randomly allocated to one half of the warm-list sample (Column 4). Accordingly, the estimated optimal targeting rule is significantly more profitable than all three benchmark policies. In conclusion, our techniques allow the fundraiser to increase net donations significantly and, hence, service and goods provision.

Third, Panel C documents that, by applying the estimated optimal targeting rule, the fundraiser would also impact outcomes besides net donations (labeled secondary outcomes). Thus, although we train our algorithm to maximize net donations, implementing the estimated rule would trigger secondary effects as a byproduct. One secondary outcome is the donation probability within the first postexperiment year (see Row C1). Specifically, by implementing the estimated optimal rule, the fundraiser would increase this probability by almost three percentage points ($5\%$) compared to the no-gift benchmark. Our proposed targeting strategy, hence, not only maximizes net donations, but also broadens the donor base compared to a scenario without gifts. In contrast to this result, the donation probability under the estimated rule is not significantly higher than under the  all-gift benchmark. Hence, although this benchmark endows many more individuals with the gift, it does not fundamentally increase the donation probability. This result suggests that intensive margin responses drive the difference in net donations between the benchmark and the estimated rule. Besides impacts on the donation probability, the table also reveals secondary effects on longer-term outcomes. For example, Row C2 of Table \ref{main_res_warm} demonstrates that, when using the outcome ``aggregate donations made within two years after the experiment,'' the positive effects of applying the estimated optimal targeting rule persists. This is an important result from the fundraiser's perspective: It highlights that the machine-learning-induced increase in net donations would not curb subsequent donations.

\begin{table}[tbp!]
	\footnotesize \singlespacing
	\begin{center}
		\caption{Out-of-sample performance of targeting rule in the cold list} \label{main_res_cold}
		\begin{tabularx}{\textwidth}{Xccccc} \toprule
			&    & Expected outcome value & \multicolumn{3}{c}{Optimal targeting vs. benchmarks}  \\
			&    & under optimal targeting  & all-gift & no-gift & random-gift \\
			\cline{3-6}   &     & (1)   & (2)   & (3)   & (4) \\ \midrule
			\multicolumn{6}{c}{\textbf{Panel A: Share of individuals that should receive the gift}} \\\midrule			
			A1. Share treated & 0.014 &  &  & &   \vspace{.15cm}\\	  \midrule		
			\multicolumn{6}{c}{\textbf{Panel B: Results for primary outcome variable}}\\\midrule
			\multicolumn{2}{l}{B1. Net donation amount}  & 0.15*** & 0.97*** & -0.005 & 0.48*** \\
			(1st year) 								     && (0.02) & (0.10) & (0.012) & (0.05) \vspace{.15cm}\\ \midrule			
			\multicolumn{6}{c}{\textbf{Panel C: Results for secondary outcome variables}} \\\midrule
			\multicolumn{2}{l}{C1. Donation probability}  & 0.009*** & -0.007*** & 0.001 & -0.003** \\
			(1st year) &       & (0.001) & (0.003) & (0.001) & (0.001) \\\\			
			\multicolumn{2}{l}{C2. Net donation amount}  & 0.44*** & 0.95*** & 0.04 & 0.50*** \\
			(1st and 2nd year) &       & (0.07) & (0.13) & (0.06) & (0.07) \\\\
			\multicolumn{2}{l}{C3. Donation probability}  & 0.017*** & -0.006* & 0.001 & -0.003 \\
			(1st and 2nd year)	&       & (0.001) & (0.003) & (0.001) & (0.002) \\\bottomrule				
		\end{tabularx}
	\end{center}
	\parbox{\textwidth}{\footnotesize{\textbf{Notes:} This table documents the out-of-sample performance of our estimated optimal targeting rule, focusing on the cold list. The goal of optimal targeting is to maximize donations, net of costs. Panel A reports the share of individuals that, according to the rule, should receive the gift. Panel B reports the expected consequences of our rule for net donations as our main outcome. Panel C, instead, focuses on secondary outcomes. Column 1 shows the expected value of the outcomes under optimal targeting. For example, we expect that, under optimal targeting, the donations, net of costs, would be $0.15$ euro. Columns 2--4 show how optimal targeting changes the outcomes relative to three benchmark scenarios: everybody receives the gift (Column 2), no one receives the gift (Column 3), and the gift is randomly assigned to half of the sample (Column 4). Donations are measured in euro. Standard errors are in parentheses. ***/**/* indicate statistical significance at the 1\%/5\%/10\% level.}}
\end{table}

\addvspace{0.1cm}\noindent\textbf{The estimated optimal targeting rule in the cold list.} Table \ref{main_res_cold} shows the out-of-sample performance of the estimated optimal targeting rule in the cold list. Due to the substantial size of the cold-list sample ($17{,}000$ individuals), all of the effects are very precisely estimated. Again, the results are very different from those for the warm list. One marked difference is that the estimated target group is much narrower in the cold list (Panel A): In line with the evidence from the sorted-effects model, the estimated optimal rule assigns the gift to just $1.4\%$ of the cold-list individuals. Given this finding, it is not surprising that the average net donation under the estimated optimal rule ($0.15$ euro) is virtually identical to that under the no-gift benchmark (Column 3 in Panel B). By contrast, the estimated optimal targeting rule outperforms the all-gift (Column 2 in Panel B) and random-gift benchmarks (Column 4 in Panel B). The reason is that campaigns that apply these two benchmarks would result in losses (due to the gifts' costs). Table \ref{main_res_cold} also presents evidence on secondary effects (Panel C). Again, we find no evidence for pull-forward or delay effects. Further, there are only minimal impacts on the donation probability. To sum up, the potential of machine-learning-based optimal targeting in the cold list is very limited, perhaps because the cold list consists of many notorious nondonors. This insight was not clear \textit{a priori} and could only be established with a flexible, data-driven approach such as ours.

\begin{table}[tbp!]
	\footnotesize \singlespacing
	\begin{center}
		\caption{Characteristics of individuals in the warm list}   \label{net_warm}
		\begin{tabular}{lccccc}\toprule
			& \multicolumn{4}{c}{Individuals targeted by the algorithm} &  \\
			& \multicolumn{2}{c}{\textbf{Yes (net donors)}} & \multicolumn{2}{c}{\textbf{No (net receivers)}} & Std. \\
			& Mean  & Std. dev. & Mean  & Std. dev. & diff. \\
			\cline{2-6}  & (1)   & (2)   & (3)   & (4)   & (5) \\\midrule
			\multicolumn{6}{c}{\textbf{Panel A: Donation history before the experiment}} \\\midrule
			A1. Num. donations prev. 8 years  & 4.097 & 2.827 & 3.900 & 2.829 & 6.934 \\
			A2. Max. donation prev. 8 years (euro)  & 39.94 & 44.61 & 34.05 & 41.89 & 13.63 \\
			A3. Total donations prev. 8 years (euro)  & 130.98 & 150.69 & 123.39 & 187.39 & 4.460 \\
			A4. Donations 1 year ago (euro)  & 22.69 & 36.49 & 19.53 & 34.60 & 8.891 \\
			A5. Donations 2 years ago (euro)  & 17.91 & 30.30 & 16.89 & 28.77 & 3.466 \\
			A6. Donations 3 years ago (euro)  & 16.61 & 27.49 & 15.62 & 27.52 & 3.593 \\
			A7. Donations 4 years ago (euro)  & 16.71 & 28.33 & 15.37 & 27.21 & 4.813 \\
			A8. Donations 5 years ago (euro)  & 15.69 & 24.48 & 15.03 & 29.96 & 2.410 \\\midrule
			\multicolumn{6}{c}{\textbf{Panel B: Geospatial information}} \\\midrule
			B1. Elevation (meters)  & 308.66 & 6.266 & 321.38 & 9.524 & 157.80 \\
			\multicolumn{6}{l}{B2. In 300 meters proximity:} \\
			\quad Number of restaurants & 10.86 & 13.30 & 6.528 & 7.711 & 39.88 \\
			\quad Number of supermarkets & 1.062 & 1.371 & 1.086 & 1.362 & 1.748 \\
			\quad Number of medical facilities & 10.17 & 13.95 & 9.298 & 12.041 & 6.703 \\
			\quad Number of cultural facilities & 0.240 & 0.796 & 0.050 & 0.241 & 32.27 \\
			\quad Number of churches & 1.166 & 1.515 & 0.934 & 1.460 & 15.60 \\
			B3. Distance to main station (km)  & 3.247 & 2.521 & 3.245 & 1.867 & 0.053 \\
			B4. Distance to city hall (km)  & 2.927 & 2.237 & 3.196 & 1.856 & 13.11 \\
			B5. Distance to main church (km)  & 2.986 & 2.365 & 3.218 & 1.836 & 10.99 \\
			B6. Distance to airport (km)  & 5.427 & 1.236 & 5.483 & 1.960 & 3.408 \\
			B7. Travel time to main station (minutes)  & 18.42 & 11.79 & 17.50 & 7.560 & 9.371 \\ \midrule
			\multicolumn{6}{c}{\textbf{Panel C: Socioeconomic characteristics}} \\\midrule
			C1. Female dummy & 0.507 &       & 0.539 &       & 6.459 \\
			C2. Single dummy  & 0.503 &       & 0.496 &       & 1.464 \\
			C3. Widowed dummy  & 0.050 &       & 0.052 &       & 0.974 \\
			C4. Age (years)  & 68.08 & 18.23 & 68.72 & 18.34 & 3.488 \\
			C5. Residency duration (years)  & 7.423 & 1.690 & 7.439 & 1.659 & 0.951 \\\midrule	
			Observations & \multicolumn{2}{c}{787} & \multicolumn{2}{c}{1,567} &  \\\bottomrule		
		\end{tabular}
	\end{center}
	\parbox{\textwidth}{\footnotesize{\textbf{Notes:} This table describes characteristics of the warm-list individuals who, according to our estimated optimal targeting rule, should receive a gift (predicted net donors) or should not receive a gift (predicted net recipients). Particularly, it reports the means and standard deviations of all observed characteristics for the predicted net donors (Columns 1--2) and the predicted net recipients (Columns 3--4). It also shows the standardized difference (Column 5). The residency duration in the urban area is censored after $8$ years. Further, we measure travel time to the main station using public transportation at 9am on weekdays. \cite{ro83} classify absolute standardized differences (std.~diff.) of more than 20 as \emph{large}.}}
\end{table}

\subsection{Characteristics of Predicted Net Donors and Net Receivers}
\label{subsec:characteristics}
A common theme in the fundraising literature is understanding the characteristics of individuals who give to charitable causes \citep{And2013}. Our machine-learning approach allows us to extend this literature by performing a broader descriptive analysis: Instead of merely describing the characteristics of givers, we can differentiate between the predicted net donors and predicted net recipients (i.e., individuals who increase donation by less than the gift's cost). The analysis, hence, reveals the characteristics of the individuals who, according to our estimated targeting rule, should receive the gift and contrasts them with the characteristics of those who should not be targeted.

\addvspace{0.1cm}\noindent\textbf{Characteristics of predicted net donors and net receivers in the warm list.}\hspace{0.2cm}
Table \ref{net_warm} focuses on the warm list. It reports the means and standard deviations of all observed characteristics for predicted net donors (Columns 1--2) and predicted net receivers (Columns 3--4). It also shows the standardized difference, a standard balance diagnostic (Column 5). The table highlights some apparent differences between the two groups. For example, before the experiment, the predicted net donors donated on average more and also more frequently than the predicted net recipients (Panel A). They also live in more central areas that are characterized by (a) lower altitudes and (b) a more lively environment with more churches, restaurants, and cultural facilities (Panel B).%
\footnote{In the urban area we study, the topology correlates with distance to the city center. Concretely, individuals living in lower altitudes live on average closer to the city hall and the main church.}
By contrast, the differences in socioeconomic characteristics are not very pronounced. If anything, predicted net donors are more frequently female (Panel C). When interpreting these results, keep in mind that the characteristics are not necessarily reflecting channels through which the impacts of the gift operate. The patterns might instead mirror effects working through correlated unobservable variables. For example, the donation history may approximate individuals' general willingness to give, and geospatial information could proxy income. In this vein, these and similar unobservable variables might channel the responses to the gift. We consider it a strength of our approach that the machine-learning algorithm can pick up the underlying forces that shape individuals' reactions to the gift without the need to collect data or explicitly model the relationships.

\addvspace{0.1cm}\noindent\textbf{Characteristics of predicted net donors and net receivers in the cold list.}\hspace{0.2cm}
Table \ref{net_cold} reports similar results for the cold list. The predicted net donors from the cold list also live in more central areas than the respective predicted net recipients, but the areas now tend to be less lively. Regarding the socioeconomic characteristics, there are more pronounced differences compared to the warm list. Relative to predicted net recipients, predicted net donors tend to have a higher likelihood of being females, singles, and newly settled residents.

\begin{table}[tbp!]
	\footnotesize \singlespacing
	\begin{center}
		\caption{Characteristics of individuals targeted by the algorithm in the cold list}   \label{net_cold}
		\begin{tabular}{lccccc}\toprule
			& \multicolumn{4}{c}{Individuals targeted by the algorithm} &  \\
			& \multicolumn{2}{c}{\textbf{Yes (net donors)}} & \multicolumn{2}{c}{\textbf{No (net receivers)}} & Std. \\
			& Mean  & Std. dev. & Mean  & Std. dev. & diff. \\
			\cline{2-6}  & (1)   & (2)   & (3)   & (4)   & (5) \\\midrule
			\multicolumn{6}{c}{\textbf{Panel A: Geospatial information}} \\\midrule
			A1. Elevation (meters)  & 313.47 & 8.183 & 316.15 & 10.34 & 28.71 \\
			\multicolumn{6}{l}{A2. In 300 meters proximity:} \\
			\quad Number of restaurants & 5.482 & 3.524 & 10.40 & 11.67 & 57.08 \\
			\quad Number of supermarkets & 0.888 & 1.122 & 1.296 & 1.502 & 30.77 \\
			\quad Number of medical facilities & 13.73 & 10.20 & 10.68 & 13.17 & 25.89 \\
			\quad Number of cultural facilities & 0.040 & 0.196 & 0.146 & 0.532 & 26.43 \\
			\quad Number of churches & 1.100 & 1.017 & 1.177 & 1.538 & 5.957 \\
			A3. Distance to main station (km)  & 1.995 & 0.890 & 2.874 & 2.028 & 56.14 \\
			A4. Distance to city hall (km)  & 1.364 & 0.645 & 2.816 & 1.885 & 103.02 \\
			A5. Distance to main church (km)  & 1.588 & 0.743 & 2.803 & 1.932 & 83.03 \\
			A6. Distance to airport (km)  & 4.143 & 1.038 & 5.567 & 1.642 & 103.68 \\
			A7. Travel time to main station (minutes)  & 12.50 & 6.312 & 16.18 & 8.680 & 48.45 \\ \midrule
			\multicolumn{6}{c}{\textbf{Panel B: Socioeconomic characteristics}} \\\midrule
			B1. Female dummy & 0.558 &       & 0.503 &       & 11.04 \\
			B2. Single dummy  & 0.713 &       & 0.642 &       & 15.24 \\
			B3. Widowed dummy  & 0.024 &       & 0.017 &       & 4.518 \\
			B4. Age (years)  & 47.58 & 21.00 & 48.41 & 19.30 & 4.132 \\
			B5. Residency duration (years)  & 5.677 & 2.964 & 5.973 & 2.818 & 10.22 \\\midrule
			Observations & \multicolumn{2}{c}{251} & \multicolumn{2}{c}{17,174} &  \\\bottomrule
		\end{tabular}
	\end{center}
	\parbox{\textwidth}{\footnotesize{\textbf{Notes:} This table describes characteristics of the cold-list individuals who, according to our estimated optimal targeting rule, should receive a gift (predicted net donors) or should not receive a gift (predicted net recipients). Particularly, it reports the means and standard deviations of all observed characteristics for the predicted net donors (Columns 1--2) and the predicted net recipients (Columns 3--4). It also shows the standardized difference (Column 5). The residency duration in the urban area is censored after $8$ years. Further, we measure travel time to the main station using public transportation at $9:00$am on weekdays. For dummy variables, the first moment is sufficient to infer the entire distribution. \cite{ro83} classify absolute standardized differences (std.~diff.) of more than 20 as \emph{large}.}}
\end{table}

\addvspace{0.1cm}\noindent\textbf{Decision trees.}\hspace{0.2cm}
A second, natural way to describe the estimated targeting rule is to plot decision trees. However, because we use a cross-validation approach to evaluate the estimated rule's out-of-sample performance, we obtain $20$ different trees per list. To reduce complexity, Figure \ref{tree} in Online Appendix \ref{app:tree}, hence, reports trees estimated using the entire sample. The first tree concerns the warm list. It shows that the previous donation amount and the elevation at the home address serve as split variables. By contrast, the second tree for the cold-list is split based on the number of restaurants near the residence and the distance to the city hall and airport. Importantly, the trees do not identify the causal effect of the donor characteristics on the net donations. Rather, they indicate correlations between donor characteristics and the gift's causal effect. Hence, we must interpret them cautiously.

\subsection{Relevant Type of Data for Optimal Targeting}
\label{subsec:datasources}
As discussed before, we estimate the optimal targeting rule using socioeconomic characteristics, donation history, and geospatial information. The next step of our analysis explores which of these data are especially powerful to estimate targeting rules. It also investigates if our algorithms require all the data to estimate effective optimal targeting rules. Besides being interesting in itself, studying this topic is vital for charities. Data collection is costly, and, frequently, some forms of data (such as socioeconomic characteristics) are unavailable. In many settings, a charity might only have access to address data before sending out written solicitations. Hence, from a charity's perspective, the usefulness and feasibility of machine-learning-based optimal targeting critically depend on the data required to target net donors effectively.

\begin{table}[tbp!]
	\footnotesize \singlespacing
	\begin{center}
		\caption{Relevant data types in the warm list}   \label{cov_warm}
		\begin{tabular}{cccccc} \toprule
			Share & Expected donations  & \multicolumn{4}{c}{Optimal targeting vs.} \\
			treated & under optimal targeting & all-gift & no-gift & random-gift & all variables \\
			(1)   & (2)   & (3)   & (4)   & (5)   & (6) \\ \midrule
			\multicolumn{6}{c}{\textbf{Panel A: Socioeconomic characteristics}}\\\midrule
			0.55 & 15.71*** & 0.24 & 0.29 & 0.27 & -1.90** \\
			& (0.79) & (0.86) & (0.77) & (0.58) & (0.89) \\\\\midrule
			\multicolumn{6}{c}{\textbf{Panel B: Donation history before the experiment}}\\\midrule
			0.12 & 17.20*** & 1.73** & 1.79** & 1.76*** & -0.41 \\
			& (0.97) & (0.82) & (0.81) & (0.58) & (0.55) \\\\\midrule
			\multicolumn{6}{c}{\textbf{Panel C: Geospatial information}}\\\midrule
			0.49 & 17.40*** & 1.92** & 1.98** & 1.95*** & -0.22 \\
			&  (0.91) & (0.84) & (0.79) & (0.58) & (0.61) \\\\\midrule
			\multicolumn{6}{c}{\textbf{Panel D: Socioeconomic characteristics and donation history}}\\\midrule
			0.11 & 17.05*** & 1.58* & 1.64** & 1.61*** & -0.56 \\
			&  (0.96) & (0.84) & (0.79) & (0.58) & (0.56) \\\\\midrule
			\multicolumn{6}{c}{\textbf{Panel E: Socioeconomic characteristics and geospatial information}}\\\midrule
			0.48 & 16.97*** & 1.50* & 1.55* & 1.53*** & -0.64 \\
			&  (0.89) & (0.84) & (0.80) & (0.58) & (0.62) \\\\\midrule
			\multicolumn{6}{c}{\textbf{Panel F: Donation history and geospatial information}}\\\midrule
			0.33 & 17.61*** & 2.14*** & 2.20*** & 2.17*** & 0 \\
			&   (0.97) & (0.82) & (0.81) & (0.58) &  \\\bottomrule
		\end{tabular}
	\end{center}
	\parbox{\textwidth}{\footnotesize{\textbf{Notes:} This table documents the out-of-sample performance of targeting rules that rely only on selected subsets of our variables, focusing on the warm list. The rule in Panel A is based only on socioeconomic characteristics, in Panel B on the donation history, in Panel C on geospatial information, in Panel D on socioeconomic characteristics and the donation history, in Panel E on socioeconomic characteristics and geospatial information, and in Panel F on the donation history and geospatial information. Column 1 shows the share of individuals who, according to the respective rule, should receive the gift. Column 2 reports expected donations under optimal targeting. Columns 3--5 show how optimal targeting changes the outcomes relative to three benchmark scenarios: the all-gift (Column 3), no-gift (Column 4), and random-gift (Column 5) benchmarks. Column 6 compares the rule that only uses the subset of variables to the rule that relies on the full set of data. Methodologically, the rules are estimated with an exact policy-learning trees and a search depth of two \citep{zho2019}. Donations are measured in euro. Standard errors are in parentheses. ***/**/* indicate statistical significance at the 1\%/5\%/10\% level.}}
\end{table}

\addvspace{0.1cm}\noindent\textbf{Relevant data in the warm list.}\hspace{0.2cm}
Table \ref{cov_warm} focuses on the warm list and explores the relevance of the different data types for the performance of the estimated optimal targeting rule. For this purpose, it evaluates the performance of several estimated targeting rules, each using only a subset of the available data. As before, the table benchmarks these more sparsely estimated rules against the all-gift (Column 3), no-gift (Column 4), and random-gift (Column 5) benchmarks. Additionally, Column 6 compares these newly estimated rules to the estimated optimal (baseline) rule obtained when using all data (reported in Table \ref{main_res_warm}).

The results are as follows: First, our baseline rule clearly outperforms a rule that relies only on socioeconomic characteristics. Also, the rule based only on socioeconomic characteristics does not outperform the three benchmarks (Panel A). As socioeconomic data are often hard to obtain, these results might not be problematic for charities. Second, our baseline rule does not significantly dominate rules that either use only data on past donations (Panel B) or use only geospatial information (Panel C). Similar to our baseline rule, these two more sparsely estimated rules also significantly outperform the three benchmarks. These findings suggest that the data on past donations and geospatial information are substitutes in targeting. Hence, charities that do not have access to details on individuals' donation histories might instead rely on publicly available geospatial data only. Third, Panels D--F further emphasize that the socioeconomic characteristics are of very limited use for machine-learning-based optimal targeting. Adding them does not improve the rules that rely solely on the donation history (Panels B and D) or geospatial information (Panels C and F). Furthermore, a rule that combines the donation history with the geospatial information performs as well as a rule that uses all the data. The reason is that our baseline rule is not relying on the socioeconomic characteristics at all.

\begin{table}[tbp!]
	\footnotesize \singlespacing
	\begin{center}
		\caption{Relevant data types in the cold list}   \label{cov_cold}
		\begin{tabular}{cccccc} \toprule
			Share & Expected donations  & \multicolumn{4}{c}{Optimal targeting vs.} \\
			treated & under optimal targeting & all-gift & no-gift & random-gift & all variables \\
			(1)   & (2)   & (3)   & (4)   & (5)   & (6) \\ \midrule
			\multicolumn{6}{c}{\textbf{Panel A: Socioeconomic characteristics}}\\\midrule
			0.015 & 0.14*** & 0.96*** & -0.014** & 0.47*** & -0.010 \\
			& (0.02) & (0.10) & (0.007) & (0.05) & (0.014) \\\\\midrule
			\multicolumn{6}{c}{\textbf{Panel B: Geospatial information}}\\\midrule
			0.014 & 0.15*** & 0.97*** & -0.005 & 0.48*** & 0 \\
			& (0.02) & (0.10) & (0.012) & (0.05) &  \\	\bottomrule
		\end{tabular} 
	\end{center}
	\parbox{\textwidth}{\footnotesize{\textbf{Notes:} This table documents the out-of-sample performance of targeting rules that rely only on selected subsets of our variables, focusing on the cold list. The rule in Panel A is based on socioeconomic characteristics only, and in Panel B on geospatial information only. Column 1 shows the share of individuals who, according to the respective rule, should receive the gift. Column 2 reports the expected donations under optimal targeting. Columns 3--5 show how optimal targeting changes the outcomes relative to three benchmark scenarios: the all-gift (Column 3), no-gift (Column 4), and random-gift (Column 5) benchmarks. Column 6 compares the rule that uses only the subset of variables to the rule that relies on the full set of data. Methodologically, the rules are estimated with exact policy-learning trees and a search depth of two \citep{zho2019}. Donations are measured in euro. Standard errors are in parentheses. ***/**/* indicate statistical significance at the 1\%/5\%/10\% level.}}
\end{table}

\addvspace{0.1cm}\noindent\textbf{Relevant data in the cold list.}\hspace{0.2cm}
Table \ref{cov_cold} reports similar analyses for the cold list. To that end, it restricts the data either to socioeconomic characteristics or geospatial information. It turns out that the socioeconomic characteristics are also redundant in the cold list: Again, our baseline targeting rule does not use socioeconomic characteristics. Thus, the results do not change when using only the geospatial information instead of all data (see Panel B).

We draw two main conclusions from this subsection. First, in the warm list, the fundraiser can significantly improve its campaigns' profits by relying only on widely available geospatial information. Put differently, the fundraiser does not necessarily need access to detailed data on past donations, and socioeconomic characteristics seem to be of little use for optimal targeting. This finding raises the attractiveness of our approach for charities that, for the sake of simplicity or due to data-collection costs, prefer to rely on a single data source. Second, in the cold list, the potential benefits of machine-learning-based optimal targeting are very limited. Given the data available, the dominant strategy is not to send the gift to individuals in the cold list.

\begin{table}[tbp!]
	\footnotesize \singlespacing
	\begin{center}
		\caption{Optimal Targeting in $2015$ based on the $2014$ rule} \label{external_validity}
		\begin{tabularx}{\textwidth}{Xccccc} \toprule
			&    & Expected outcome value & \multicolumn{3}{c}{Optimal targeting vs. benchmarks}  \\
			&    & under optimal targeting  & all-gift & no-gift & random-gift \\
			\cline{3-6}   &     & (1)   & (2)   & (3)   & (4) \\ \midrule
			\multicolumn{6}{c}{\textbf{Panel A: Share of individuals that should receive the gift}} \\\midrule			
			A1. Share treated & 	0.356 &  &  & &    \\	  
			(matched sample)  & 	      &  &  & &    \vspace{.15cm}\\
			A1. Share treated & 	0.353 &  &  & &    \\	  
			(full 2015 sample)  & 	      &  &  & &    \vspace{.15cm}\\\midrule	
			\multicolumn{6}{c}{\textbf{Panel B: Results for primary outcome variable}}\\\midrule
			\multicolumn{2}{l}{B1. Net donation amount}  & 	19.85*** &	2.64* &	
			2.30**	& 2.47** \\
			(matched sample) 								     &&	(1.80)	& (1.57) &	(1.10) &	(0.96) \vspace{.15cm}\\
			\multicolumn{2}{l}{B2. Net donation amount}  & 	17.16*** &	-0.88 &	
			0.45	& -0.22 \\
			(full $2015$ sample) 								     &&	(0.74)	& (0.81) & (0.53) &	(0.48) \vspace{.15cm}\\\bottomrule	%
		\end{tabularx}
	\end{center}
	\parbox{\textwidth}{\footnotesize{\textbf{Notes:} This table documents the simulated performance of our $2014$ optimal targeting rule when the charity would implemented it in $2015$. Panel A reports the share of individuals that, according to the rule, should receive the gift. Panel B reports the expected consequences of our rule for net donations as our main outcome. Column 1 shows the expected value of the outcomes under optimal targeting. For example, we expect that, under optimal targeting, the donations, net of costs, would be $17.01$ euro. Columns 2--4 show how optimal targeting changes the outcomes relative to three benchmark scenarios: everybody receives the gift (Column 2), no one receives the gift (Column 3), and the gift is randomly assigned to half of the sample (Column 4). Donations are measured in euro. Standard errors in parentheses. ***/**/* indicate statistical significance at the 1\%/5\%/10\% level.}}
\end{table}

\subsection{External Validity of the Estimated Optimal Targeting Rule} \label{subsec:stability}
This subsection explores if the estimated optimal targeting rule is externally valid across two identical consecutive fundraising campaigns. To that end, we test whether the charity can increase net donations in the second follow-up campaign by distributing gifts according to the rule estimated on first-campaign data. We consider two dimensions of external validity, which we label ``time stability'' and ``sample stability.'' According to our definition, an estimated optimal targeting rule is time stable if the charity can beneficially apply it to a follow-up campaign run on a \textit{similar} sample of potential donors. Instead, a rule that satisfies sample stability can even be successfully applied to consecutive campaigns run on samples with \textit{different characteristics}.

\addvspace{0.1cm}\noindent\textbf{Follow-up experiment.}\hspace{0.2cm} 
We study the estimated rule's external validity by exploiting data from a follow-up experiment that took place in a subsequent year using different participants. Specifically, in $2015$, we randomly allocated a new sample of $3,616$ warm-list individuals to the control group and the gift treatment (treatment probability: $50\%$).%
\footnote{Given the evidence from the $2014$ campaign, we decided to consider only warm-list individuals. In Table \ref{tabH1} of Online Appendix \ref{app:2015}, we report the balance of the observed characteristics by treatment status.} 
Notably, the $2015$ sample vastly differs from the $2014$ sample in its observable characteristics. For example, individuals in the $2015$ sample donated more, and more frequently before the experiment, than those in the $2014$ sample (see Table \ref{tabH2} in Online Appendix \ref{app:2015}). Their response to the gift was also more pronounced: The average effect of the gift on net donations was $1.33$ euro in $2015$ compared to $0.08$ in $2014$. 

\addvspace{0.1cm}\noindent\textbf{Studying both dimensions of external validity.}\hspace{0.2cm} 
The follow-up experiment allows us to study both dimensions of external validity. Because the $2015$ and $2014$ samples differ, we can test sample stability by examining how the rule estimated on $2014$ data (hereinafter, $2014$ rule) performs in the full $2015$ sample. A test of time stability, instead, requires similar samples in both years. To construct comparable samples, we employ a caliper propensity score matching approach with a radius of $0.1$ that matches the $2015$ sample to the $2014$ sample (see Appendix \ref{app:external_validity} for details). We then evaluate time stability by assessing the $2014$ rule's performance in the matched $2015$ sample.

\addvspace{0.1cm}\noindent\textbf{Results of our external-validity checks.}\hspace{0.2cm} 
Table \ref{external_validity} demonstrates how net donations would have changed in $2015$ when the charity would have allocated the gift following the $2014$ rule. Again, it presents results relative to our three benchmarks. Furthermore, it separately considers the matched sample and the full $2015$ sample. The results are as follows: First, the $2014$ rule suggests that $35.6\%$ of the warm-list individuals in the matched sample and $35.3\%$ in the full $2015$ sample should receive the gift. Second, the $2014$ rule seems to be time stable. The charity would have benefited substantially from applying the $2014$ rule to the matched $2015$ sample. Hereby, it could have increased net donations by $2.64$ euro relative to the all-gift benchmark (Column 2 in Row B1) and $2.30$ euro relative to the no-gift benchmark (Column 3 in Row B1). These effects are comparable to those in $2014$ (see Table \ref{main_res_warm}). Third, however, we cannot confirm sample stability when considering the full $2015$ sample. By applying the $2014$ rule to this sample, the charity would have been unable to increase net donations relative to the all-gift and no-gift benchmarks.%
\footnote{The coefficient for the all-gift benchmark is negative because, in $2015$, the gift significantly increased average net donations. Hence, it is difficult to outperform the all-gift benchmark.}
This result is not surprising: If the sample changes too much from $2014$ to $2015$, the second-year data contains few individuals similar to those on which the algorithm was trained. This likely affects the performance of the optimal targeting rule in $2015$, at least in the presence of effect heterogeneity (individuals in the $2014$ and $2015$ samples respond differently).

\addvspace{0.1cm}\noindent\textbf{Varying the sample differences.}\hspace{0.2cm} 
Given the previous discussion, one occurring question is how much the charity can change the sample such that the optimal targeting rule is still effective. Online Appendix \ref{app:external_validity} provides evidence on this topic. Specifically, it explores how similar the $2014$ and $2015$ samples must be for the $2014$ rule to still increase net donations in $2015$. To that end, we sequentially increase the caliper radius, resulting in the $2015$ sample differing more and more strongly from the $2014$ sample. Figure \ref{warm_2015} in Online Appendix \ref{app:external_validity} presents the results. One message of the figure is that the $2014$ rule significantly outperforms the no-gift benchmark up to a caliper radius of $0.25$ (i.e., for quite different samples). By contrast, because the gift increases average net donations in the full sample, it is much more difficult to outperform the all-gift benchmark. For caliper radii above $0.1$, the effects are no longer statistically significant (but still positive in expectation).

To sum up, in our context, the estimated optimal targeting rule is time stable over one year; it is also stable over different samples, but only if they do not differ too much. Thus, when using previously estimated rules, our charity needs to hold both samples as similar as possible to obtain the best result (e.g., by using an appropriate sampling design). Even though this finding is not surprising, it never has been shown before. 

\subsection{Sensitivity Analysis}
\label{subsec:robustness}
Our final step is to investigate the robustness of the estimated optimal targeting rules to different estimation approaches. First, we consider alternative depths of the exact policy-learning tree (one and three).%
\footnote{In the cold list, exact policy-learning trees of ``depth three'' are infeasible due to computational constraints.}
Second, we compare the results of the exact policy-learning trees to the results of standard CARTs \citep{brei84}. Third, for the CARTs, we also consider one version in which we use cross-validation to select the tree depth in a data-driven way.%
\footnote{We use the Gini index for tree splitting and a $10$-fold cross-validation procedure to select the optimal tree depth. In the warm list, the number of terminal leaves varies between four and thirteen across the 20 different cross-validated trees, with an average of $5$. In the cold list, the number of terminal leaves varies between one and nineteen across the $20$ different cross-validated trees, with an average of $3.6$.}
Fourth, we employ weighted logit as a standard estimator. Fifth, we use two alternative machine-learning estimators: logit lasso \citep[][]{hast16} and classification forests \citep{brei01}.%
\footnote{We build $1{,}000$ trees for the classification forest. We draw a $50\%$ random subsample with replacement for each tree and randomly select $50\%$ of the baseline characteristics. We use the Gini index for tree splitting. We restrict the minimum size of the terminal leaf to $50$ observations.}
For the logit estimator, we consider two different model specifications. The baseline specification includes all observed characteristics linearly ($24$ variables in the warm list and $16$ variables in the cold list). The flexible specification additionally includes squared terms of continuous variables and first-order interactions between most characteristics ($320$ variables in the warm list and $148$ variables in the cold list). The logit-lasso method selects the relevant characteristics from the flexible model specification.%
\footnote{We specify the penalty $\lambda$ of the logit lasso that minimizes the misclassification error using a $10$-fold cross-validation approach.}

Table \ref{robust_warm} in Online Appendix \ref{app:sensitivity} reports the results of the sensitivity analysis for the warm list, and Table \ref{robust_cold} focuses on the cold list. In the warm list, our baseline optimal targeting strategy clearly dominates the alternative specifications: The baseline optimal targeting rules of all the alternative estimators yield lower net donations than our main specification. Specifically, the exact policy-learning tree with depth one and the logit with the flexible model specification have the lowest out-of-sample performance. The CART with the cross-validated tree depth is the only alternative estimator that also significantly outperforms the all-gift and no-gift benchmark allocation rules. For the cold list, the logit-lasso specification and the CART with cross-validated tree depth yield $0.01$ euro higher net donations than our main specification. However, our main finding that the fundraiser should not send the gift to cold-list individuals persists.

\section{Conclusion}
\label{sec:conclusion}
This paper studies machine-learning-based optimal targeting of fundraising instruments by exploiting data from a natural field experiment. The underlying idea of optimal targeting is that fundraisers can maximize a campaign's profits by directing a fundraising instrument to individuals who increase their donations in response to the instrument by more than its cost. We label those individuals \emph{net donors}. However, charities do not observe the set of net donors. We employ a machine-learning algorithm to estimate the relationship between individual characteristics and the potential donors' response to small unconditional gifts. Based on this algorithm, we can predict the subset of net donors and, hence, estimate machine-learning-based optimal targeting rules.

Our paper's first key message is that, in the warm list, machine-learning-based optimal targeting substantially boosts the charity's net donations. In our application, net donations increase by about $14\%$ compared to the all-gift and no-gift benchmarks. Notably, the benefits of machine-learning-based optimal targeting even materialize when (a) applying the estimated optimal targeting rule to a later campaign conducted in a similar sample or (b) relying only on widely available geospatial data. Hence, charities can easily apply the proposed strategies to raise additional funds, net of costs. The second message is that, in the cold list, the approach does not raise donations sufficiently to cover the fundraising instrument's additional costs. We conclude that the fundraiser should not target cold-list individuals at all. We also document that the increase in net donations stems from heterogeneities in donors' responses to fundraising activities. Previous literature has suggested that such heterogeneities exist, for example, by providing mixed evidence on the effectiveness of unconditional gifts on giving \citep{Fal2007,yin2020coins,ALP2008}.

In conclusion, our paper demonstrates that machine-learning-based optimal targeting can significantly increase the cost effectiveness of fundraising. One particularly noteworthy benefit of our applied machine-learning toolkit is that it allows charities to target fundraising efforts agnostically in a wide variety of contexts. Thus, to optimize targeting, charities do not need to develop a theoretical foundation or make strong assumptions on the functional relationship between individual characteristics and giving. We are, therefore, confident that the proposed approach offers an accessible way forward to improve the effectiveness of fundraising. We are also looking forward to evolving research applying similar techniques to alternative fundraising instruments and different settings. It will also be interesting to see how our findings generalize across these alternative tools and environments.

\bibliographystyle{ecca}
\bibliography{bib_2021_03_06_clean}

\clearpage

\renewcommand\appendix{\par
	\setcounter{section}{0}%
	\setcounter{subsection}{0}%
	\setcounter{table}{0}%
	\renewcommand\thesection{\Alph{section}}%
	\renewcommand\thetable{\Alph{section}.\arabic{table}}}
\renewcommand\thefigure{\Alph{section}.\arabic{figure}}

\clearpage
\setcounter{page}{1}
\setcounter{footnote}{0}
\setcounter{equation}{0}
\setcounter{footnote}{0}
\renewcommand*{\thefootnote}{\fnsymbol{footnote}}

\begin{center}
	{\huge Online Appendix to 
		
		``Optimal Targeting in Fundraising''} \bigskip
	
	{\Large {Tobias Cagala, Ulrich Glogowsky, \\Johannes Rincke, and Anthony Strittmatter\footnote{Cagala: Deutsche Bundesbank (tobias.cagala@bundesbank.de); Glogowsky: Johannes Kepler University Linz and CESifo (ulrich.glogowsky@jku.at); Rincke: University of Erlangen-Nuremberg (johannes.rincke@fau.de); Strittmatter: CREST-ENSAE, Institut Polytechnique Paris and CESifo (anthony.strittmatter@ensae.fr).} }} \bigskip
\end{center}

\subsection*{\Large Sections:}
\begin{enumerate}
	\item[A.] Descriptives and Balance of Observables
	\item[B.] Identification of ATEs and CATEs
	\item[C.] Identification with AIPW Scores
	\item[D.] Nuisance Parameters
	\item[E.] Additional Heterogeneity Results
	\item[F.] Example Decision Trees
	\item[G.] Descriptives for the 2015 Sample
	\item[H.] External Validity of the 2014 Rule
	\item[I.] Sensitivity Analysis
\end{enumerate}
\newpage

\begin{appendix}
	\setcounter{footnote}{0}
	\renewcommand*{\thefootnote}{\arabic{footnote}}
	
	\numberwithin{equation}{section}
	\setcounter{table}{0}
	
	\section{Descriptives and Balance of Observables}
	\label{app:descriptives}

    \begin{table}[tbph!]
    	\small \singlespacing
    	\begin{center}
    		\caption{Descriptive statistics of donation amount and donation probability} \label{desc_out}
    		\begin{tabularx}{\textwidth}{Xcccccc}\toprule
    			& Mean  & Std. dev. & Skewn. & Kurt. & Min.  & Max. \\
    			\cline{2-7}     & (1)   & (2)   & (3)   & (4    & (5)   & (6) \\\midrule
    			\multicolumn{7}{c}{\textbf{Panel A: Warm list}} \\\hline
    			\multicolumn{7}{l}{1st year after the experiment} \\
    			\quad A1. Donation amount (euro) & 16.02 & 30.39 & 4.69 & 36.02 & 0     & 450 \\
    			\quad A2. Donation dummy & 0.49  & &   &   & 0     & 1 \\ \\
    			\multicolumn{7}{l}{1st and 2nd year after the experiment} \\
    			\quad A3. Donation amount (euro) & 30.48 & 53.19 & 4.96  & 46.18 & 0     & 900 \\
    			\quad A4. Donation dummy & 0.57  &  &   &   & 0     & 1 \\\midrule
    			\multicolumn{7}{c}{\textbf{Panel B: Cold list}} \\\midrule
    			\multicolumn{7}{l}{1st year after the experiment} \\
    			\quad  B1. Donation amount (euro) & 0.18  & 3.00  & 38.38 & 2,046.1  & 0     & 200 \\
    			\quad B2. Donation dummy & 0.009  &   &  &  & 0     & 1 \\ \\
    			\multicolumn{7}{l}{1st and 2nd year after the experiment} \\
    			\quad  B3. Donation amount (euro) & 0.43 & 4.85& 23.12  & 776.5 & 0     & 240 \\
    			\quad B4. Donation dummy & 0.017  &   &   &   & 0     & 1 \\\bottomrule
    		\end{tabularx}
    	\end{center}
    	\parbox{\textwidth}{\footnotesize{\textbf{Notes:} This table shows descriptive statistics for our primary outcomes, (a) the donation amount and (b) variables indicating whether a person donated. 
    	Further, it tracks our outcomes over two periods, the first year after and the first two years after the experiment. The table considers the warm list (Panel A) and cold list (Panel B) separately. 
    	}}
    \end{table}	
    \clearpage

	\begin{table}[tbph!]
		\small  \singlespacing
		\begin{center}
			\caption{Means and standard deviations of observable characteristics}\label{desc}
			\begin{tabularx}{\textwidth}{Xccccc}\toprule
				& \multicolumn{2}{c}{Warm list} &&  \multicolumn{2}{c}{Cold list} \\
				& Mean  & Std. dev.  && Mean  & Std. dev.  \\
				\cline{2-3}\cline{5-6}     & (1)   & (2)   & & (3)   & (4)     \\\midrule
				\multicolumn{6}{c}{\textbf{Panel A. Socioeconomic characteristics}} \\\midrule
				Female dummy  & 0.53  &    &&  0.50  &    \\
				Single dummy & 0.50  &    &&  0.64  &    \\
				Widowed dummy & 0.05  &   &&  0.02  &    \\
				Age (years) & 68.51 & 18.30 &&  48.40 & 19.32  \\
				Residency duration in urban area (years) & 7.43  & 1.67     && 5.97  & 2.82   \\\midrule
				\multicolumn{6}{c}{\textbf{Panel B: Donation history before the experiment}} \\\midrule
				Number of donations previous 8 years & 3.97  & 2.83 &&0&  \\
				Max. donations previous 8 years (euro) & 36.02 & 42.90 &&0& \\
				Total donations previous 8 years (euro) & 125.9 & 176.0 &&0& \\
				Donations 1 year ago (euro) & 20.59 & 35.27 &&0& \\
				Donations 2 years ago (euro) & 17.23 & 29.29 &&0& \\
				Donations 3 years ago (euro) & 15.95 & 27.51 &&0& \\
				Donations 4 years ago (euro) & 15.82 & 27.59 &&0& \\
				Donations 5 years ago (euro) & 15.25 & 28.24 &&0& \\\midrule
				\multicolumn{6}{c}{\textbf{Panel C: Geospatial information about home address}} \\\midrule
				Elevation (meters) & 317.1 & 10.46  &&  316.1 & 10.32  \\
				\multicolumn{6}{l}{In 300 meters proximity:} \\
				\quad Number of restaurants  & 7.98  & 10.14  &  & 10.33 & 11.61  \\
				\quad Number of supermarkets  & 1.08  & 1.36   && 1.29  & 1.50   \\
				\quad Number of medical facilities  & 9.59  & 12.72  && 10.72 & 13.13  \\
				\quad Number of cultural facilities  & 0.11  & 0.51   & & 0.14  & 0.53   \\
				\quad Number of churches  & 1.01  & 1.48   && 1.18  & 1.53   \\
				Distance to main station (km) & 3.25  & 2.11  && 2.86  & 2.02   \\
				Distance to city hall (km) & 3.11  & 2.00   && 2.79  & 1.88   \\
				Distance to main church (km) & 3.14  & 2.03   && 2.79  & 1.93    \\
				Distance to airport (km) & 5.46  & 1.75   && 5.55  & 1.64      \\
				Travel time to main station (minutes) & 17.81 & 9.20   && 16.13 & 8.66   \\\midrule
				Observations & \multicolumn{2}{c}{2,354} && \multicolumn{2}{c}{17,425} \\\bottomrule
			\end{tabularx}
		\end{center}
		\parbox{\textwidth}{\footnotesize{\textbf{Notes:} This table describes the characteristics of cold and warm-list individuals. The donation history in the cold list is zero, because we only measure the donations to the specific fundraiser we cooperate with. The residency duration in the urban area is censored after 8 years. We measure travel time to the main station using public transportation at $9:00$am on weekdays. For dummy variables, the first moment is sufficient to infer the entire distribution. \cite{ro83} classify absolute standardized difference (std. diff.) of more than 20 as \emph{large}.}}
	\end{table}
	\clearpage
	
	\begin{table}[tbph!]
		\small  \singlespacing
		\begin{center}
			\caption{Balance of observable characteristics in the warm list} \label{bal_warm}
			\begin{tabularx}{\textwidth}{Xccccc}\toprule
				& \multicolumn{2}{c}{Treatment group} & \multicolumn{2}{c}{Control group} & Std. \\
				& Mean  & Std. dev. & Mean  & Std. dev. & diff. \\
				\cline{2-6}     & (1)   & (2)   & (3)   & (4)   & (5) \\\midrule
				\multicolumn{6}{c}{\textbf{Panel A: Socioeconomic characteristics}} \\\midrule
				Female dummy  & 0.53  &   & 0.53  &  & 1.22 \\
				Single dummy & 0.50  &   & 0.49  &   & 2.04 \\
				Widowed dummy & 0.05  &   & 0.06  &   & 3.98 \\
				Age (years) & 68.57 & 18.31 & 68.45 & 18.31 & 0.69 \\
				Residency duration (years) & 7.41  & 1.70  & 7.46  & 1.63  & 3.13 \\\midrule
				\multicolumn{6}{c}{\textbf{Panel B: Donation history before the experiment}} \\\midrule
				Num. donations prev. 8 years & 3.94  & 2.81  & 3.99  & 2.85  & 1.98 \\
				Max. don. prev. 8 years (euro) & 36.49 & 46.63 & 35.54 & 38.81 & 2.22 \\
				Total don. prev. 8 years (euro) & 126.64 & 181.52 & 125.21 & 170.31 & 0.82 \\
				Donations 1 year ago (euro) & 21.19 & 39.73 & 19.98 & 30.13 & 3.41 \\
				Donations 2 years ago (euro) & 17.28 & 29.55 & 17.18 & 29.03 & 0.32 \\
				Donations 3 years ago (euro) & 16.11 & 28.13 & 15.80 & 26.88 & 1.13 \\
				Donations 4 years ago (euro) & 16.30 & 28.79 & 15.34 & 26.34 & 3.48 \\
				Donations 5 years ago (euro) & 14.83 & 28.53 & 15.67 & 27.95 & 2.96 \\\midrule
				\multicolumn{6}{c}{\textbf{Panel C: Geospatial information about home address}} \\\midrule
				Elevation (meters) & 317.4 & 10.58 & 316.9 & 10.35 & 4.89 \\
				\multicolumn{6}{l}{In 300 meters proximity:} \\
				\quad Number of restaurants  & 7.86  & 10.15 & 8.10  & 10.14 & 2.42 \\
				\quad Number of supermarkets  & 1.11  & 1.37  & 1.05  & 1.36  & 4.83 \\
				\quad Number of medical facilities  & 9.52  & 12.52 & 9.66  & 12.91 & 1.03 \\
				\quad Number of cultural facilities  & 0.11  & 0.52  & 0.11  & 0.50  & 0.22 \\
				\quad Number of churches  & 1.03  & 1.50  & 1.00  & 1.46  & 1.83 \\
				Distance to main station (km) & 3.24  & 2.04  & 3.25  & 2.18  & 0.59 \\
				Distance to city hall (km) & 3.09  & 1.93  & 3.12  & 2.06  & 1.15 \\
				Distance to main church (km) & 3.13  & 1.96  & 3.15  & 2.10  & 0.87 \\
				Distance to airport (km) & 5.44  & 1.77  & 5.49  & 1.74  & 2.47 \\
				Travel time to main station (minutes) & 17.72 & 8.80  & 17.90 & 9.58  & 1.94 \\\midrule
				Observations & \multicolumn{2}{c}{1,180} & \multicolumn{2}{c}{1,174} &  \\\bottomrule
			\end{tabularx}
		\end{center}
		\parbox{\textwidth}{\footnotesize{\textbf{Notes:} This table describes the characteristics of the warm-list individuals in the treatment and control groups. The residency duration in the urban area is censored after 8 years. We measure travel time to the main station using public transportation at $9:00$am on weekdays. For dummy variables, the first moment is sufficient to infer the entire distribution. \cite{ro83} classify absolute standardized difference (std. diff.) of more than 20 as \emph{large}.}}
	\end{table}
	\clearpage
	
	\begin{table}[tbph!]
		\small \singlespacing
		\begin{center}
			\caption{Balance of observable characteristics in the cold list} \label{bal_cold}
			\begin{tabularx}{\textwidth}{Xccccc}\toprule
				& \multicolumn{2}{c}{Treatment group} & \multicolumn{2}{c}{Control group} & Std. \\
				& Mean  & Std. dev. & Mean  & Std. dev. & diff. \\
				\cline{2-6}      & (1)   & (2)   & (3)   & (4)   & (5) \\\midrule
				\multicolumn{6}{c}{\textbf{Panel A: Socioeconomic characteristics}} \\\midrule
				Female dummy  & 0.50  &   & 0.50  &   & 0.14 \\
				Single dummy & 0.64  &  & 0.64  &   & 0.23 \\
				Widowed dummy & 0.02  &  & 0.02  &   & 0.35 \\
				Age (years) & 48.34 & 19.25 & 48.41 & 19.33 & 0.34 \\
				Residency duration (years) & 5.96  & 2.82  & 5.97  & 2.82 & 0.36 \\\midrule
				\multicolumn{6}{c}{\textbf{Panel B: Geospatial information about home address}} \\\midrule
				Elevation (meters) & 316.2 & 10.42 & 316.1 & 10.31 & 0.40 \\
				\multicolumn{6}{l}{In 300 meters proximity:} \\
				\quad Number of restaurants  & 10.02 & 11.29 & 10.38 & 11.66 & 3.17 \\
				\quad Number of supermarkets  & 1.31  & 1.50  & 1.29  & 1.50  & 1.14 \\
				\quad Number of medical facilities  & 10.40 & 12.91 & 10.77 & 13.17 & 2.85 \\
				\quad Number of cultural facilities  & 0.14  & 0.51  & 0.15  & 0.53  & 1.10 \\
				\quad Number of churches  & 1.13  & 1.49  & 1.18  & 1.54  & 3.27 \\
				Distance to main station (km) & 2.89  & 2.02  & 2.86  & 2.02  & 1.69 \\
				Distance to city hall (km) & 2.81  & 1.88  & 2.79  & 1.88  & 0.90 \\
				Distance to main church (km) & 2.81  & 1.93  & 2.78  & 1.93  & 1.19 \\
				Distance to airport (km) & 5.55  & 1.65  & 5.55  & 1.64  & 0.07 \\
				Travel time to main station (minutes) & 16.25 & 8.80  & 16.11 & 8.64  & 1.64 \\\midrule
				Observations & \multicolumn{2}{c}{2,283} & \multicolumn{2}{c}{15,142} &  \\\bottomrule
			\end{tabularx}
		\end{center}
		\parbox{\textwidth}{\footnotesize{\textbf{Notes:} This table describes the characteristics of the cold-list individuals in the treatment and control groups. The residency duration in the urban area is censored after 8 years. We measure travel time to the main station using public transportation at $9:00$am on weekdays. For dummy variables, the first moment is sufficient to infer the entire distribution. \cite{ro83} classify absolute standardized difference (std. diff.) of more than 20 as \emph{large}. }}
	\end{table}
	\clearpage
	
	\section{Identification of ATEs and CATEs} \setcounter{table}{0}
	\label{app:CATEs}
	To achieve identification, we have to assume that the stratified randomization was appropriately conducted, ensuring $p(z,x)= Pr(D_i= 1|Z_i=z, X_i=x)= Pr(D_i= 1|Z_i=z)= p(z)$ and $(Y_i(1),Y_i(-1)) \ci D_i|Z_i=z$. Furthermore, we have to assume that the probability of receiving the gift is between zero and one, $0< p(z) <1$.
	
	Under these assumptions, the following equations prove that the CATEs are identified from observable data:
	\begin{align*}
	\delta(x)
	& \stackrel{\text{\tiny LIE}}{=} E_{Z|X=x} [E[Y_{i}(1)-Y_{i}(-1)|Z_i=z,X_{i}=x]],\\
	& \!\!\!\!\!\!\stackrel{\text{\tiny Experiment}}{=} E_{Z|X=x} [E[Y_{i}(1)|D_i=1,Z_i=z, X_{i}=x] - E[Y_{i}(-1)|D_i=-1, Z_i=z,X_{i}=x]],\\
	& \!\!\stackrel{\text{\tiny SUTVA}}{=} E_{Z|X=x} [E[Y_{i}|D_i=1,Z_i=z, X_{i}=x] - E[Y_{i}|D_i=-1, Z_i=z,X_{i}=x]],
	\end{align*}
	where the first equality is an application of the law of iterative expectations (LIE), the second equality holds under the experimental design, and the last equality under the SUTVA. The identification proof for the ATEs follows from the LIE, $\delta = E[\delta(X_i)]$.
	\clearpage
	
	\section{Identification with AIPW Scores} \setcounter{table}{0}
	\label{app:AIPW}
	To prove that $\delta = E[{\Gamma}_i]$ and $\delta(x)= E[{\Gamma}_i|X_i=x]$, it is sufficient to prove that
	\begin{align*}
	E\left[ \left.  Y_{i}(1)  \right| X_i=x \right] =& E\left[ \left.  {\Gamma}_i(1)  \right| X_i=x \right]= E \left[ \left.  {\mu}_1(Z_{i}) + \frac{1+D_{i}}{2} \cdot \frac{Y_{i} - {\mu}_{1}(Z_{i}) }{{p}(Z_{i})}   \right| X_i=x \right],\\
	E\left[ \left.  Y_{i}(-1)  \right| X_i=x \right] =& E\left[ \left.  {\Gamma}_i(-1)  \right| X_i=x \right] = E \left[ \left.  {\mu}_{-1}(Z_{i}) - \frac{D_{i}-1}{2} \cdot\frac{Y_{i} - {\mu}_{-1}(Z_{i}) }{1-{p}(Z_{i})} \right| X_i=x \right].
	\end{align*}
	We focus on $E[{\Gamma}_i(1)|X_i=x]$. It follows:
	\begin{align}
	E\left[ \left.  {\Gamma}_i(1)  \right| X_i=x \right] =& E \left[ \left.  {\mu}_1(Z_{i})  + \frac{1+D_{i}}{2} \cdot \frac{Y_{i} - {\mu}_{1}(Z_{i}) }{{p}(Z_{i})} \right| X_i=x \right],\label{eq:Equality 1}\\
	=& E \left[ \left.  \frac{1+D_{i}}{2} \cdot \frac{Y_{i}  }{{p}(Z_{i})} \right| X_i=x \right] +  E \left[ \left. \left(  {p}(Z_{i})  - \frac{1+D_{i}}{2} \right) \cdot \frac{{\mu}_{1}(Z_{i}) }{{p}(Z_{i})} \right| X_i=x \right],\label{eq:Equality 2}\\
	=& E_{Z|X=x} \left[ E\left[ \left.  \frac{1+D_{i}}{2} \cdot \frac{Y_{i}  }{{p}(Z_i)} \right| X_i=x, Z_i=z \right] \right] \notag\\
	& +  E_{Z|X=x} \left[ E\left[\left. \left(  {p}(Z_i)  - \frac{1+D_{i}}{2} \right) \cdot \frac{{\mu}_{1}(Z_i) }{{p}(Z_i)} \right| X_i=x, Z_i=z \right] \right],\label{eq:Equality 3}\\
	=& E_{Z|X=x} \left[ E\left[ \left.  \frac{1+D_{i}}{2} \cdot \frac{Y_{i}  }{{p}(z,x)} \right| X_i=x, Z_i=z \right]\right] \notag\\
	& + \underbrace{ E_{Z|X=x} \left[ E\left[ \left. \left(  {p}(z,x)  - \frac{1+D_{i}}{2} \right) \cdot \frac{{\mu}_{1}(Z_i) }{{p}(z,x)} \right| X_i=x, Z_i=z \right] \right]}_{=0},\label{eq:Equality 4}\\
	=& E_{Z|X=x} E\left[ \left[ \left.    \frac{1\{D_i=1\} Y_{i}  }{{p}(z,x)} \right| X_i=x, Z_i=z \right]\right],\label{eq:Equality 5}\\
	=& E_{Z|X=x}E\left[  \left[ \left.     Y_{i}   \right| D_i=1, X_i=x, Z_i=z \right]\right],\label{eq:Equality 6}\\
	=& E_{Z|X=x} E\left[ \left[ \left.    Y_{i} (1)  \right| D_i=1, X_i=x, Z_i=z \right]\right],\label{eq:Equality 7}\\
	=& E_{Z|X=x}E\left[  \left[ \left.   Y_{i} (1)  \right|  X_i=x, Z_i=z \right]\right],\label{eq:Equality 8}\\
	=& E \left[ \left.     Y_{i} (1)  \right|  X_i=x \right].\label{eq:Equality 9}
	\end{align}
	In \eqref{eq:Equality 1}, we use the definition of ${\Gamma}_i(1)$. In \eqref{eq:Equality 2}, we rearrange terms. In \eqref{eq:Equality 3}, we apply the law of iterative expectations. In \eqref{eq:Equality 4}, we exploit that $p(z)=p(z,x)$ because of our experimental design (only the characteristics in $Z_i$ have an impact on the probability of receiving the gift). Note that the second right-side term cancels, because $p(z,x)= E \left[ \left. (1+D_{i})/2 \right| X_i=x, Z_i=z \right]$. In \eqref{eq:Equality 5}, we replace $(1+D_{i})/2$ with the indicator function $1\{D_i=1\}$. In \eqref{eq:Equality 6}, we apply the discrete law of iterative expectations backwards. In \eqref{eq:Equality 7} and \eqref{eq:Equality 8}, we use $(Y_i(1),Y_i(-1)) \ci D_i|Z_i=z$, the conditional-independence assumption, which holds by the experimental design. In \eqref{eq:Equality 9}, we apply the law of iterative expectations backwards. This finishes the proof that $E\left[ \left.  Y_{i}(1) \right| X_i=x \right]= E[{\Gamma}_i(1)|X_i=x]$. The proof that $E\left[ \left.  Y_{i}(-1) \right| X_i=x \right]= E[{\Gamma}_i(-1)|X_i=x]$ proceeds analogous.
	\clearpage
	
	\section{Nuisance Parameters} \setcounter{table}{0}
	\label{app:nuisance}
	\begin{table}[tbph!]
		\small  \singlespacing
		\begin{center}
			\caption{Coefficients of nuisance parameters} \label{nuisance}
			\begin{tabularx}{\textwidth}{Xcccccc}
				\toprule
				& \multicolumn{2}{c}{Logit} & \multicolumn{4}{c}{OLS} \\
				& \multicolumn{2}{c}{Donation } & \multicolumn{2}{c}{Donations} & \multicolumn{2}{c}{Donations } \\
				& \multicolumn{2}{c}{ dummy} & \multicolumn{2}{c}{ without gift} & \multicolumn{2}{c}{ with gift} \\
				\cline{2-7}   & \multicolumn{2}{c}{(1)} & \multicolumn{2}{c}{(2)} & \multicolumn{2}{c}{(3)} \\
				& Coef. & Std. err. & Coef. & Std. err. & Coef. & Std. err. \\\midrule
				\multicolumn{7}{c}{\textbf{Panel A: Warm list}} \\\midrule
				Female dummy & -0.009 & 0.023 & -3.67** & 1.69  & -4.99** & 1.91 \\
				Single dummy & 0.009 & 0.023 & 0.31  & 1.74  & 1.22  & 1.98 \\
				Widowed dummy & -0.041 & 0.050 & -1.29 & 3.54  & 2.40  & 4.41 \\
				Age quintiles: &       &       &       &       &       &  \\
				\quad 2nd quintile & 0.005 & 0.029 & -3.96* & 2.13  & 1.22  & 2.42 \\
				\quad 3rd quintile & 0.004 & 0.030 & -5.68** & 2.20  & -3.20 & 2.51 \\
				\quad 4th quintile & 0.006 & 0.031 & -2.61 & 2.31  & 1.12  & 2.64 \\
				\multicolumn{7}{l}{Baseline willingness to donate quintiles:} \\
				\quad 2nd quintile & -0.002 & 0.031 & -29.64*** & 2.33  & -25.76*** & 2.65 \\
				\quad 3rd quintile & 0.001 & 0.030 & -24.81*** & 2.23  & -19.02*** & 2.55 \\
				\quad 4th quintile & 0.003 & 0.030 & -12.73*** & 2.23  & -7.94*** & 2.54 \\
				Intercept & 0.498*** & 0.031 & 37.13*** & 2.27  & 32.44*** & 2.59 \\\midrule
				\multicolumn{7}{c}{\textbf{Panel B: Cold list}} \\\midrule
				Female dummy & -0.003 & 0.046 & -0.002 & 0.04  & 0.18  & 0.20 \\
				Single dummy & -0.003 & 0.052 & -0.10** & 0.05  & -0.28 & 0.23 \\
				Widowed dummy & 0.027 & 0.176 & -0.01 & 0.17  & 0.93  & 0.77 \\
				Age quintiles: &       &       &       &       &       &  \\
				\quad 2nd quintile & 0.009 & 0.065 & 0.09  & 0.06  & 0.22  & 0.28 \\
				\quad 3rd quintile & 0.002 & 0.067 & 0.10  & 0.06  & 0.03  & 0.29 \\
				\quad 4th quintile & 0.001 & 0.068 & 0.20*** & 0.06  & 0.45  & 0.30 \\
				Intercept & -1.891*** & 0.068 & 0.14** & 0.07  & 0.32  & 0.30 \\ \bottomrule
			\end{tabularx}
		\end{center}
		\parbox{\textwidth}{\footnotesize{\textbf{Notes:} This table shows the coefficients of the nuisance parameters. We additionally controlled for a dummy indicating whether the church sent a flyer in the year before the experiment. Donations (euro) are measured during the first year after the gift was sent. ***/**/* indicate statistical significance at the 1\%/5\%/10\% level. }} 
	\end{table}
	\clearpage
	
	\section{Additional Heterogeneity Results} \setcounter{table}{0}
	\label{app:sorted}
	\begin{table}[tbph!]
		\begin{center}
			\caption{Mean characteristics of the groups with the 10\% largest and smallest treatment effects in the warm list} \label{high_low_10_warm} \footnotesize \singlespacing
			\begin{tabularx}{\textwidth}{Xccccccc} \toprule
				& \multicolumn{2}{c}{10\% Largest } & \multicolumn{2}{c}{10\% Smallest } & \multicolumn{3}{c}{Difference} \\
				& Mean  & Std. err. & Mean  & Std. err. & Mean  & Std. err. & $JP$-val.  \\
				\cline{2-8}       & (1)   & (2)   & (3)   & (4)   & (5)   & (6)   & (7)    \\\midrule
				\multicolumn{8}{c}{\textbf{Panel A: Socioeconomic characteristics}}  \\ \midrule
			Female dummy   & 0.49  & 0.04  & 0.41  & 0.05  & 0.08  & 0.07  & 1.00 \\
    				Single dummy  & 0.56  & 0.05  & 0.49  & 0.05  & 0.07  & 0.08  & 1.00 \\
    				Widowed dummy  & 0.11  & 0.03  & 0.05  & 0.03  & 0.06  & 0.04  & 0.96 \\
    				Age (years)  & 70.81 & 1.88  & 67.55 & 1.94  & 3.26  & 3.13  & 1.00 \\
    				Residency duration (years)  & 7.29  & 0.21  & 6.94  & 0.24  & 0.35  & 0.34  & 1.00 \\
    				  \midrule
				\multicolumn{8}{c}{\textbf{Panel B: Donation history before the experiment} }\\\midrule
			Num. donations prev. 8 years  & 3.63  & 0.25  & 4.45  & 0.28  & -0.82 & 0.41  & 0.71 \\
    				Max. don. prev. 8 years (euro)  & 55.31 & 7.05  & 70.04 & 8.49  & -14.73 & 13.25 & 1.00 \\
    				Total don. prev. 8 years (euro)  & 161.02 & 23.72 & 271.08 & 30.73 & -110.06 & 42.48 & 0.30 \\
    				Donations 1 year ago (euro)  & 25.30 & 5.12  & 51.05 & 7.31  & -25.76 & 10.63 & 0.37 \\
    				Donations 2 years ago (euro)  & 19.41 & 4.47  & 41.86 & 5.39  & -22.45 & 7.64  & 0.15 \\
    				Donations 3 years ago (euro)  & 26.16 & 3.63  & 36.37 & 4.44  & -10.21 & 6.22  & 0.90 \\
    				Donations 4 years ago (euro)  & 23.61 & 3.98  & 29.62 & 4.55  & -6.01 & 6.63  & 1.00 \\
    				Donations 5 years ago (euro)  & 21.32 & 4.35  & 32.15 & 4.96  & -10.83 & 7.37  & 0.95 \\
    				\midrule
				\multicolumn{8}{c}{\textbf{Panel C: Geospatial information about home address}} \\\midrule
		    				Elevation (meters)  & 315.43 & 1.25  & 321.34 & 1.48  & -5.91 & 2.16  & 0.23 \\
    				\multicolumn{8}{l}{In 300 meters proximity:}  \\
    				\quad Number of restaurants   & 8.96  & 1.33  & 7.48  & 1.15  & 1.48  & 2.13  & 1.00 \\
    				\quad Number of supermarkets   & 0.79  & 0.17  & 1.12  & 0.15  & -0.33 & 0.25  & 0.99 \\
    				\quad Number of medical facilities   & 9.32  & 1.63  & 9.36  & 1.39  & -0.03 & 2.40  & 1.00 \\
    				\quad Number of cultural facilities   & 0.35  & 0.08  & 0.24  & 0.07  & 0.11  & 0.12  & 1.00 \\
    			\quad Number of churches   & 1.15  & 0.18  & 1.31  & 0.17  & -0.16 & 0.32  & 1.00 \\
    				Distance to main station (km)  & 3.70  & 0.26  & 3.48  & 0.25  & 0.22  & 0.39  & 1.00 \\
    				Distance to city hall (km)  & 3.57  & 0.25  & 3.30  & 0.25  & 0.28  & 0.38  & 1.00 \\
    				Distance to main church (km)  & 3.60  & 0.26  & 3.36  & 0.26  & 0.23  & 0.38  & 1.00 \\
    				Distance to airport (km)  & 5.86  & 0.17  & 5.38  & 0.15  & 0.48  & 0.25  & 0.75 \\
    				Time to main station (minutes)  & 19.77 & 1.17  & 17.97 & 1.09  & 1.79  & 1.70  & 1.00 \\
         \bottomrule
			\end{tabularx}
		\end{center}
		\parbox{\textwidth}{\footnotesize{\textbf{Notes:} This table shows the mean characteristics of the groups with the 10\% largest and smallest treatment effects in the warm list. We report joint $p$-values ($JP$-value), which account for simultaneous inference on several characteristics. We employ the so-called \emph{single-step} methods to control the family-wise error rate \citep[see, e.g.,][for details]{sort}. Standard errors are calculated with a multiplier bootstrap using 500 replications. The residency duration in the urban area is censored after 8 years. We measure travel time to the main station using public transportation at $9:00$am on weekdays.}}
	\end{table}
	\clearpage 
	
	\begin{table}[tbph!]
		\begin{center}
			\caption{Mean characteristics of the groups with the 10\% largest and smallest treatment effects in the cold list} \label{high_low_10_cold} \footnotesize \singlespacing
			\begin{tabularx}{\textwidth}{Xccccccc} \toprule
				& \multicolumn{2}{c}{10\% Largest } & \multicolumn{2}{c}{10\% Smallest } & \multicolumn{3}{c}{Difference} \\
				& Mean  & Std. err. & Mean  & Std. err. & Mean  & Std. err. & $JP$-val.  \\
				\cline{2-8}       & (1)   & (2)   & (3)   & (4)   & (5)   & (6)   & (7)    \\\midrule
				\multicolumn{8}{c}{\textbf{Panel A: Socioeconomic characteristics}}  \\\midrule
			Female dummy   & 0.71  & 0.08  & 0.50  & 0.05  & 0.21  & 0.12  & 0.74 \\
    				Single dummy  & 0.52  & 0.11  & 0.62  & 0.05  & -0.10 & 0.12  & 1.00 \\
    				Widowed dummy  & 0.11  & 0.03  & 0.06  & 0.01  & 0.05  & 0.03  & 0.75 \\
    				Age (years)  & 55.76 & 3.75  & 51.85 & 1.65  & 3.91  & 4.06  & 1.00 \\
    				Residency duration (years)  & 5.63  & 0.65  & 6.56  & 0.22  & -0.93 & 0.72  & 0.94 \\
    			\midrule
				\multicolumn{8}{c}{\textbf{Panel B: Geospatial information about home address}} \\\midrule
				Elevation (meters) & 320.25 & 1.36  & 318.75 & 1.04  & 1.50  & 1.53  & 0.99 \\
    				
				\multicolumn{8}{l}{In 300 meters proximity:}\\
	    				\quad Number of restaurants   & 10.16 & 1.67  & 10.10 & 1.13  & 0.06  & 1.77  & 1.00 \\
    				\quad Number of supermarkets   & 1.35  & 0.18  & 1.21  & 0.18  & 0.13  & 0.25  & 1.00 \\
    				\quad Number of medical facilities   & 18.77 & 2.84  & 11.49 & 1.77  & 7.28  & 2.84  & 0.26 \\
    				\quad Number of cultural facilities  & 0.09  & 0.06  & 0.39  & 0.06  & -0.30 & 0.08  & 0.04 \\
    				\quad Number of churches   & 1.13  & 0.31  & 1.42  & 0.14  & -0.29 & 0.34  & 1.00 \\
    				Distance to main station (km)  & 3.33  & 0.31  & 3.87  & 0.24  & -0.54 & 0.31  & 0.77 \\
    				Distance to city hall (km)  & 2.62  & 0.35  & 3.64  & 0.25  & -1.02 & 0.34  & 0.10 \\
    				Distance to main church (km)  & 2.90  & 0.32  & 3.70  & 0.24  & -0.80 & 0.32  & 0.28 \\
    				Distance to airport (km)  & 4.24  & 0.24  & 5.42  & 0.16  & -1.18 & 0.34  & 0.05 \\
    				Time to main station (minutes)  & 16.48 & 1.12  & 20.18 & 1.05  & -3.70 & 1.24  & 0.11 \\	\bottomrule
			\end{tabularx}
		\end{center}
		\parbox{\textwidth}{\footnotesize{\textbf{Notes:} This table shows the mean characteristics of the groups with the 10\% largest and smallest treatment effects in the cold list. We report joint $p$-values ($JP$-value), which account for simultaneous inference on several characteristics. We employ the so-called \emph{single-step} methods to control the family-wise error rate \citep[see, e.g.,][for details]{sort}. Standard errors are calculated with a multiplier bootstrap using 500 replications. The residency duration in the urban area is censored after 8 years. We measure travel time to the main station using public transportation at $9:00$am on weekdays.  }}
	\end{table}
	\clearpage

\begin{table}[tbph!]
\begin{center}
\caption{Heterogeneity test: Best linear prediction method} \label{het_test} \small  \singlespacing
\begin{tabular}{lcc}  \toprule
	&Warm List & Cold List \\
\cline{2-3} &(1) & (2) \\	
	\midrule
Average effects         &0.995 &0.960**\\
                         & (1.052) & (0.522)\\
                         &[0.172] &[0.033]\\
Heterogeneity loadings &0.959* &0.511\\
                        &(0.710) &(1.069)\\
                         &[0.088] &[0.316]\\ 	\bottomrule
\end{tabular}
\end{center}
	\parbox{\textwidth}{\footnotesize{\textbf{Notes:} We use a causal forest to implement the best linear prediction method \citep[see][for details]{Chernozhukov2020,ath19b}. Standard errors are in parentheses and p-values for the hypothesis that the parameter is not positive are in brackets. Positive and statistically significant coefficients of the heterogeneity loadings provide evidence for detectable effect heterogeneity based on the observed characteristics. ***/**/* indicate statistical significance at the 1\%/5\%/10\% level.  }}
\end{table}

		\clearpage
		
	\section{Example Decision Trees} \setcounter{figure}{0}
	\label{app:tree}
	
	\begin{figure}[tbph!]\begin{center}
			\caption{Illustration of decision tree} \label{tree}
			\begin{tabular}{c}
				\includegraphics[width=.95 \textwidth]{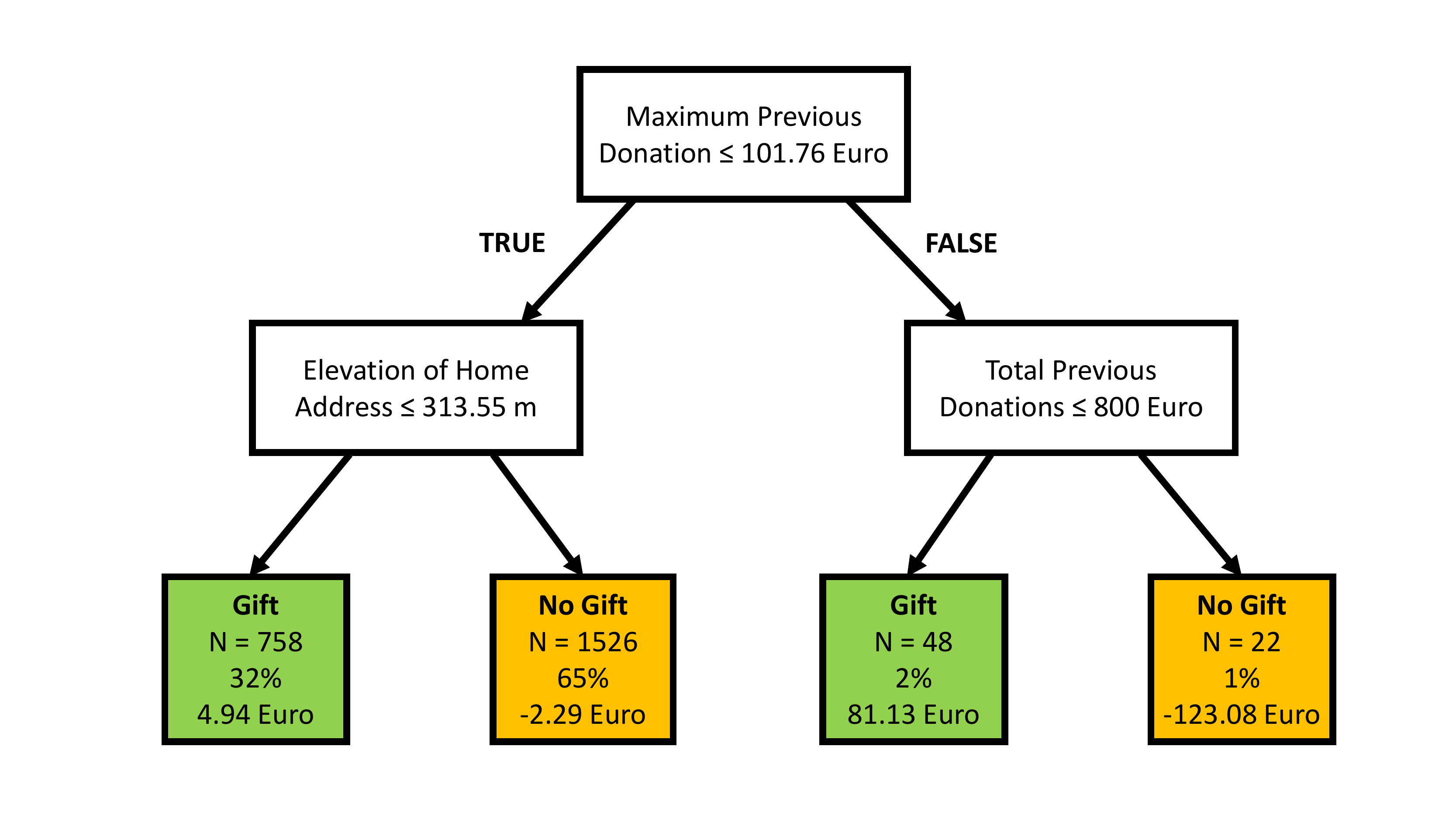} \\
				a) Warm list\\
				\includegraphics[width=.95 \textwidth]{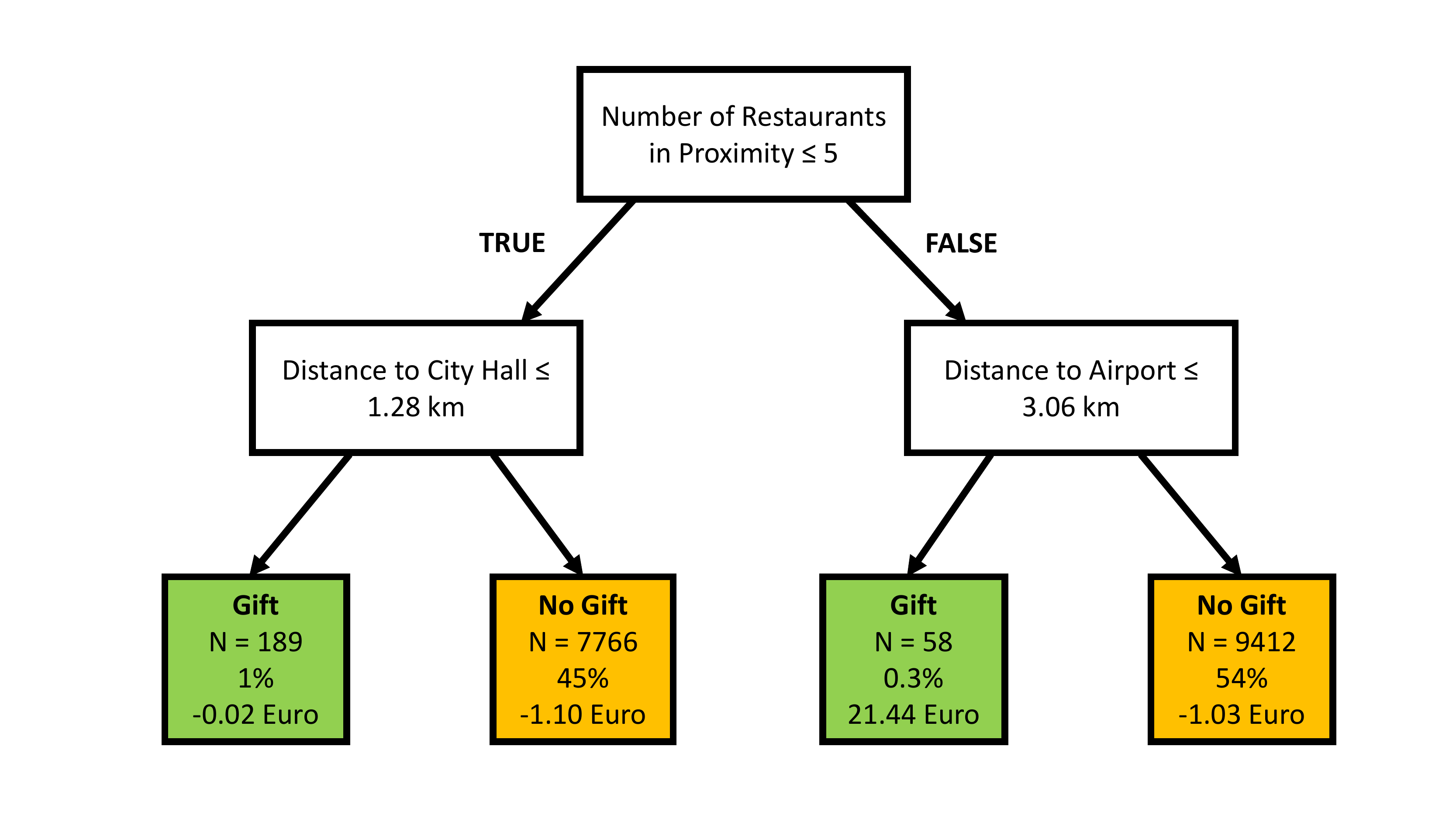}\\
				b) Cold list \\
			\end{tabular}
			
			\parbox{\textwidth}{\footnotesize{\textbf{Notes:} This figure illustrates one example decision tree. For each terminal leaf, it reports the total and relative number of observations and the effect of the gift on net donations during the first year after the experiment. To derive the decision trees, we estimate exact policy-learning trees with a search depth of two \citep{zho2019}.}}
		\end{center}
	\end{figure}
	\clearpage

	\clearpage
    \section{Descriptives for the 2015 Sample} \setcounter{table}{0}
	\label{app:2015}
	\vspace{-.5cm}
	
	\begin{table}[tbph!]
		\small  \singlespacing
		\begin{center}
			\caption{Balance of observable characteristics in the warm list 2015} \label{tabH1}
			\begin{tabularx}{\textwidth}{Xccccc}\toprule
				& \multicolumn{2}{c}{Treatment group} & \multicolumn{2}{c}{Control group} & Std. \\
				& Mean  & Std. dev. & Mean  & Std. dev. & diff. \\
				\cline{2-6}     & (1)   & (2)   & (3)   & (4)   & (5) \\\midrule
				\multicolumn{6}{c}{\textbf{Panel A: Socioeconomic characteristics}} \\\midrule
				Female dummy   & 0.52  &    & 0.52  &    & 0 \\
				Single dummy  & 0.47  &    & 0.48  &   & 2.11 \\
				Widowed dummy & 0.05  &   & 0.05  &  & 0.76 \\
				Age (years)  & 70.17 & 17.26 & 70.25 & 16.71 & 0.48 \\
				Residency duration (years) & 7.71  & 1.03  & 7.8   & 0.85  & 8.99 \\ \midrule
				\multicolumn{6}{c}{\textbf{Panel B: Donation history before the experiment}} \\\midrule
				Num. donations prev. 8 years & 4.52  & 2.81  & 4.55  & 2.8   & 0.85 \\
				Max. don. prev. 8 years (euro) & 36.66 & 45.1  & 34.68 & 38.04 & 4.74 \\
				Total don. prev. 8 years (euro)& 147.64 & 212.32 & 140.11 & 191.29 & 3.73 \\
				Donations 1 year ago (euro)  & 18.3  & 30.26 & 17.96 & 30.92 & 1.1 \\
				Donations 2 years ago (euro) & 20.71 & 33.06 & 19.28 & 32.17 & 4.4 \\
				Donations 3 years ago (euro) & 21.12 & 33.62 & 20.12 & 32.22 & 3.02 \\
				Donations 4 years ago (euro) & 20.71 & 35.86 & 19.58 & 30.89 & 3.39 \\
				Donations 5 years ago (euro) & 18.15 & 32.89 & 17.97 & 29.38 & 0.59 \\\midrule
				\multicolumn{6}{c}{\textbf{Panel C: Geospatial information about home address}} \\\midrule
				Elevation (meters) & 317.01 & 10.37 & 316.8 & 10.75 & 1.92 \\
				\multicolumn{6}{l}{In 300 meters proximity:} \\
				\quad Number of restaurants   & 8.11  & 11.07 & 7.54  & 10.4  & 5.28 \\
				\quad Number of supermarkets & 1.11  & 1.47  & 1.06  & 1.38  & 3.61 \\
				\quad Number of medical facilities  & 9.52  & 12.27 & 8.87  & 11.83 & 5.33 \\
				\quad Number of cultural facilities  & 0.16  & 0.73  & 0.13  & 0.65  & 4.09 \\
				\quad Number of churches & 1.11  & 1.63  & 1.1   & 1.63  & 0.71 \\
				Distance to main station (km) & 3.28  & 2.21  & 3.36  & 2.19  & 3.46 \\
				Distance to city hall (km) & 3.16  & 2.07  & 3.2   & 2.06  & 2.25 \\
				Distance to main church (km) & 3.19  & 2.12  & 3.25  & 2.11  & 2.78 \\
				Distance to airport (km)  & 5.43  & 1.75  & 5.41  & 1.77  & 1.16 \\
				Travel time to main station (minutes) & 17.53 & 9.93  & 17.71 & 9.69  & 1.78 \\\midrule
				Observations & \multicolumn{2}{c}{1,806} & \multicolumn{2}{c}{1,806} &  \\\bottomrule
			\end{tabularx}
		\end{center}
		\parbox{\textwidth}{\footnotesize{\textbf{Notes:} This table describes the characteristics of the 2015 warm-list individuals in the treatment and control groups. The residency duration in the urban area is censored after 8 years. We measure travel time to the main station using public transportation at $9:00$am on weekdays. For dummy variables, the first moment is sufficient to infer the entire distribution. \cite{ro83} classify absolute standardized difference (std. diff.) of more than 20 as \emph{large}.}}
	\end{table}

	\clearpage
	
		\begin{table}[tbph!]
		\small  \singlespacing
		\begin{center}
			\caption{Comparing observable characteristics in the 2014 and 2015 warm lists samples} \label{tabH2}
			\begin{tabularx}{\textwidth}{Xccccc}\toprule
				& \multicolumn{2}{c}{2014 sample} & \multicolumn{2}{c}{2015 sample} & Std. \\
				& Mean  & Std. dev. & Mean  & Std. dev. & diff. \\
				\cline{2-6}     & (1)   & (2)   & (3)   & (4)   & (5) \\\midrule
					\multicolumn{6}{c}{\textbf{Panel A: Outcome variables}} \\\midrule
							Donation amount (euro)  & 16.02 & 30.39 & 17.96 & 31.51 & 6.26 \\
				Donation dummy & 0.49  &   & 0.58  &  & 17.94 \\\midrule
				\multicolumn{6}{c}{\textbf{Panel B: Socioeconomic characteristics}} \\\midrule
				Female dummy  & 0.53  &    & 0.52  &    & 0.93 \\
				Single dummy & 0.5   &    & 0.47  &    & 5.48 \\
				Widowed dummy & 0.05  &   & 0.05  &   & 0.4 \\
				Age (years) & 68.51 & 18.3  & 70.21 & 16.99 & 9.63 \\
				Residency duration (years) & 7.43  & 1.67  & 7.75  & 0.95  & 23.6 \\\midrule
				\multicolumn{6}{c}{\textbf{Panel C: Donation history before the experiment}} \\\midrule
				Num. donations prev. 8 years & 3.97  & 2.83  & 4.54  & 2.81  & 20.26 \\
				Max. don. prev. 8 years (euro)  & 36.02 & 42.9  & 35.67 & 41.72 & 0.82 \\
				Total don. prev. 8 years (euro) & 125.93 & 175.98 & 143.87 & 202.09 & 9.47 \\
				Donations 1 year ago (euro) & 20.59 & 35.27 & 18.13 & 30.59 & 7.44 \\
				Donations 2 years ago (euro)  & 17.23 & 29.29 & 19.99 & 32.62 & 8.91 \\
				Donations 3 years ago (euro) & 15.95 & 27.51 & 20.62 & 32.92 & 15.38 \\
				Donations 4 years ago (euro)  & 15.82 & 27.59 & 20.14 & 33.47 & 14.09 \\
				Donations 5 years ago (euro)  & 15.25 & 28.24 & 18.06 & 31.18 & 9.45 \\\midrule
				\multicolumn{6}{c}{\textbf{Panel D: Geospatial information about home address}} \\\midrule
				Elevation (meters) & 317.13 & 10.46 & 316.91 & 10.56 & 2.13 \\
				\multicolumn{6}{l}{In 300 meters proximity:} \\
				\quad Number of restaurants  & 7.98  & 10.14 & 7.83  & 10.75 & 1.45 \\
				\quad Number of supermarkets  & 1.08  & 1.36  & 1.09  & 1.43  & 0.57 \\
				\quad Number of medical facilities & 9.59  & 12.72 & 9.2   & 12.06 & 3.18 \\
				\quad Number of cultural facilities   & 0.11  & 0.51  & 0.15  & 0.69  & 5.38 \\
				\quad Number of churches & 1.01  & 1.48  & 1.1   & 1.63  & 5.79 \\
				Distance to main station (km) & 3.25  & 2.11  & 3.32  & 2.2   & 3.41 \\
				Distance to city hall (km) & 3.11  & 2.00     & 3.18  & 2.07  & 3.61 \\
				Distance to main church (km) & 3.14  & 2.03  & 3.22  & 2.12  & 3.79 \\
				Distance to airport (km)  & 5.46  & 1.75  & 5.42  & 1.76  & 2.62 \\
				Travel time to main station (minutes)& 17.81 & 9.2   & 17.62 & 9.81  & 1.96 \\\midrule
				Observations & \multicolumn{2}{c}{2,354} & \multicolumn{2}{c}{3,612} &  \\\bottomrule
			\end{tabularx}
		\end{center}
		\parbox{\textwidth}{\footnotesize{\textbf{Notes:} This table describes the characteristics of the warm-list individuals between the 2014 and 2015 samples. The residency duration in the urban area is censored after 8 years. We measure travel time to the main station using public transportation at $9:00$am on weekdays. For dummy variables, the first moment is sufficient to infer the entire distribution. \cite{ro83} classify absolute standardized difference (std. diff.) of more than 20 as \emph{large}.}}
	\end{table}

    \clearpage

	\section{External Validity of the 2014 Rule} \setcounter{table}{0} \setcounter{figure}{0}
	\label{app:external_validity}
	
\begin{figure}[tbph!]
    	\begin{center}
    		\caption{Optimal targeting in $2015$ based on the $2014$ rule} \label{warm_2015}
    			\includegraphics[width=\textwidth]{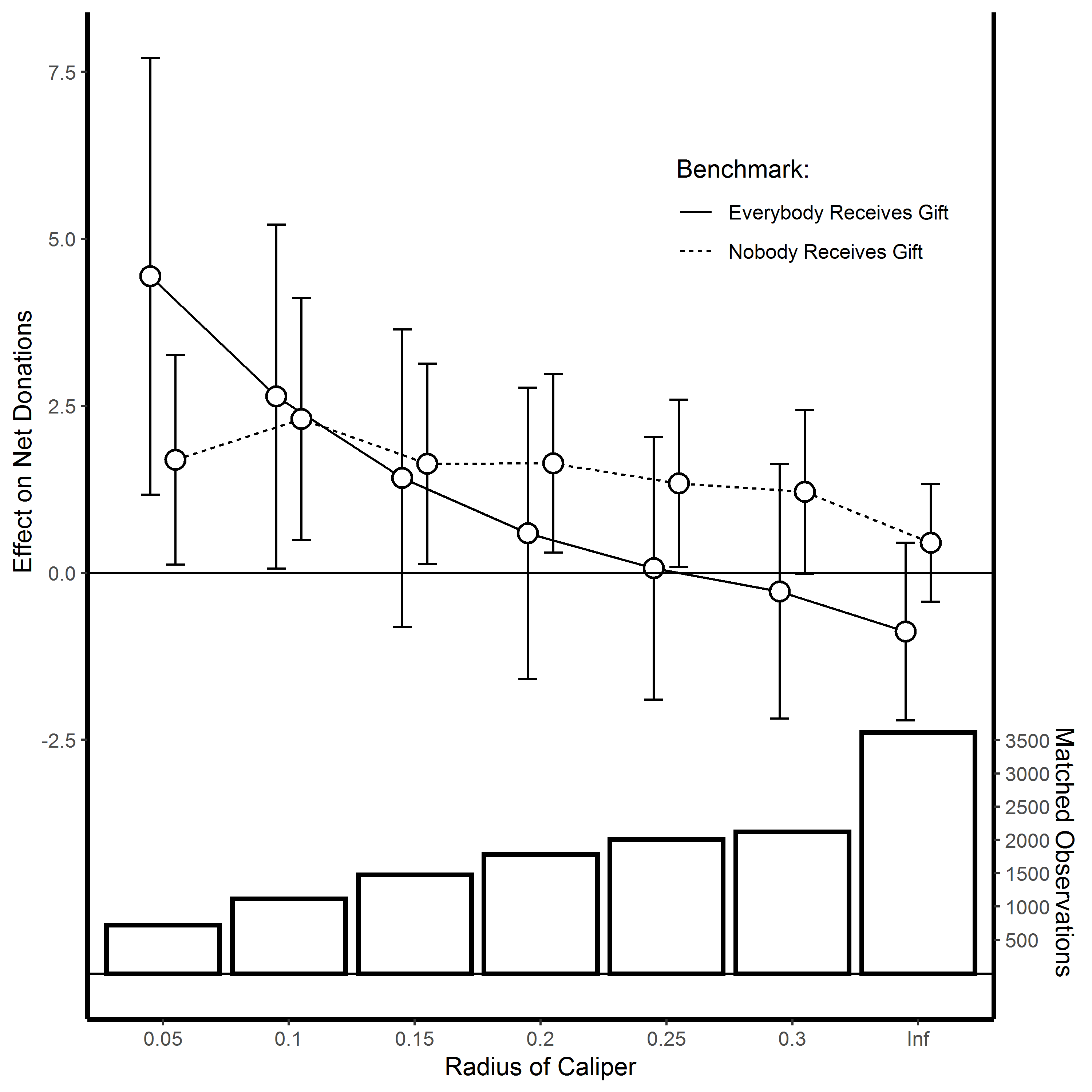}
    		\parbox{\textwidth}{\footnotesize{\textbf{Notes:} This figure shows the average effect on net donations when we apply the 2014 targeting rule to new data from 2015. We vary the radius of the caliper matching. A small radius implies that the matched 2015 sample is comparable to the 2014 sample (and \textit{vice versa}). The radius Inf indicates that we use the entire 2015 sample ($N= 3,616$). The capped bars report 90\% confidence intervals.}}
    	\end{center}
    \end{figure}

	\noindent\textbf{Caliper propensity score matching:} To test time stability, we examine how the $2014$ rule performs in a comparable sample in $2015$. As mentioned, the $2014$ and $2015$ samples are different in their characteristics. Hence, we need to construct matching samples for our time-stability test. For that purpose, we employ a standard caliper propensity matching approach. The details of our approach are as follows: First, we merge the $2014$ and $2015$ samples and estimate, for each observation, the conditional probability that it belongs to the $2014$ sample (using a random forest). Second, we match an observation from the $2015$ sample to each observation in the $2014$ sample using a nearest neighbor matching approach. As a distance metric, we utilize the absolute difference in the conditional probability to belong to the $2014$ sample. Third, we drop observations from the matched sample whenever the closest absolute difference in the conditional probability to belong to the $2014$ sample lies above a certain threshold (radius). A smaller radius implies that the matched $2015$ sample is more comparable to the $2014$ sample (and \textit{vice versa}). As a standard radius, we use a value of $0.1$. Figure \ref{warm_2015}, additionally, shows the robustness of our results to the radius width.
	\clearpage

    \clearpage

	\section{Sensitivity Analysis} \setcounter{table}{0}
	\label{app:sensitivity}
	
	\begin{table}[tbph!]
		\begin{center}
			\small  \singlespacing
			\caption{Results for alternative estimators in the warm list}   \label{robust_warm}
			\begin{tabular}{lcccc} \toprule
				& Share & Net  & \multicolumn{2}{c}{Optimal targeting vs.} \\
				& treated & donations & all-gift & no-gift \\
				\cline{2-5}    & (1)   & (2)   & (3)   & (4) \\\midrule
				\textbf{Logit} &       &       &       &  \\
				\quad   Baseline model & 0.47 & 16.19*** & 0.72 & 0.78 \\
				&       & (0.90) & (0.81) & (0.82) \\
				\quad   Flexible model & 0.42 & 16.28*** & 0.81 & 0.86 \\
				&       & (0.87) & (0.91) & (0.71) \\
				\textbf{Logit lasso} & 0.83 & 15.86*** & 0.39 & 0.45 \\
				&       & (0.91) & (0.62) & (0.98) \\
				\textbf{Exact policy-learning tree} &       &       &       &  \\
				\quad  depth = 1 & 0.39 & 15.05*** & -0.42 & -0.36 \\
				&       & (0.77) & (0.98) & (0.61) \\
				\quad depth = 3 &  0.34  & 15.49***       &  0.02     & 0.08 \\
				&       &  (0.88)     &  (0.87)     & (0.76)  \\
				\textbf{CART}  &       &       &       &  \\
				\quad depth = 2 & 0.11 & 16.10*** & 0.63 & 0.68 \\
				&       & (0.94) & (0.93) & (0.69) \\
				\quad  Cross-validated depth & 0.33 & 17.40*** & 1.93** & 1.98** \\
				&       & (0.96) & (0.83) & (0.80) \\
				\textbf{Classification forest} & 0.41 & 15.89*** & 0.41 & 0.47 \\
				&       & (0.83) & (0.89) & (0.73) \\\bottomrule
			\end{tabular}
		\end{center}
		\parbox{\textwidth}{\footnotesize{\textbf{Notes:} This table shows the results of our robustness checks for warm-list individuals. The outcome variable is the donation amount (euro) during the first year after the gift was sent. Standard errors are in parentheses. ***/**/* indicate statistical significance at the 1\%/5\%/10\% level. }}
	\end{table}
	\clearpage
	
	\begin{table}[tbph!]
		\begin{center}
			\small  \singlespacing
			\caption{Results for alternative estimators in the cold list}   \label{robust_cold}
			\begin{tabular}{lcccc} \toprule
				& Share & Net  & \multicolumn{2}{c}{Optimal targeting vs.} \\
				& treated & donations & all-gift & no-gift \\
				\cline{2-5}    & (1)   & (2)   & (3)   & (4) \\\midrule
				\textbf{Logit} &       &       &       &  \\
				\quad Baseline model & 0.047 & 0.15*** & 0.96*** & -0.01 \\
				&       & (0.04) & (0.10) & (0.03) \\
				\quad Flexible model & 0.054 & 0.10*** & 0.92*** & -0.06* \\
				&       & (0.02) & (0.10) & (0.01) \\
				\textbf{Logit lasso} & 0.0003 & 0.15*** & 0.97*** & -0.002 \\
				&       & (0.02) & (0.10) & (0.001) \\
				\textbf{Exact policy-learning tree} &       &       &       &  \\
				\quad depth = 1 & 0.071 & 0.06*** & 0.87*** & -0.10*** \\
				&       & (0.02) & (0.10) & (0.01) \\
				\textbf{CART}  &       &       &       &  \\
				\quad depth = 2 & 0.042 & 0.10*** & 0.92*** & -0.06*** \\
				&       & (0.02) & (0.10) & (0.02) \\
				\quad Cross-validated depth & 0.001 & 0.16*** & 0.97*** & -0.001** \\
				&       & (0.02) & (0.10) & (0.0003) \\
				\textbf{Classification forest} & 0.001 & 0.15*** & 0.97*** & -0.005 \\
				&       & (0.02) & (0.10) & (0.003) \\\bottomrule
			\end{tabular}
		\end{center}
		\parbox{\textwidth}{\footnotesize{\textbf{Notes:} This table shows the results of our robustness checks for cold-list individuals. The outcome variable is the donation amount (euro) during the first year after the gift was sent. In the cold list, implementing an exact policy-learning tree with a depth of three is computationally infeasible. Standard errors are in parentheses. ***/**/* indicate statistical significance at the 1\%/5\%/10\% level. }}
	\end{table}
    \clearpage
    
		\begin{table}[tbph!]
		\begin{center}
			\small  \singlespacing
			\caption{Results for alternative specification of nuisance parameters: \\ Population propensity score}   \label{pop_weight}
			\begin{tabular}{lcccc} \toprule
				& Share & Net  & \multicolumn{2}{c}{Optimal targeting vs.} \\
				& treated & donations & all-gift & no-gift \\
				\cline{2-5}    & (1)   & (2)   & (3)   & (4) \\
				\midrule		
			\multicolumn{5}{c}{\textbf{Panel A: Warm list}}\\\midrule
			
		Net donation amount & 0.334 & 17.63*** &	2.15***&	2.21***
 \\
		(1st year)		&       & (0.97)&	(0.82)&	(0.82)
\\
							\midrule		
			\multicolumn{5}{c}{\textbf{Panel B: Cold list}}\\\midrule
					Net donation amount & 0.014 & 0.15***&	0.97***&	-0.005
 \\
			(1st year)	&       & (0.02)&	(0.10)&	(0.012)
 \\
		\bottomrule
			\end{tabular}
		\end{center}
		\parbox{\textwidth}{\footnotesize{\textbf{Notes:} This table replicates the main results of Tables \ref{main_res_warm} and \ref{main_res_cold} when using the population propensity score instead of the estimated propensity score. Standard errors are in parentheses. ***/**/* indicate statistical significance at the 1\%/5\%/10\% level. }}
	\end{table}
	\clearpage				

			\begin{table}[tbph!]
		\begin{center}
			\small  \singlespacing
			\caption{Results for alternative specification of nuisance parameters: \\ Cross-fitted lasso}   \label{lasso_nuisance}
			\begin{tabular}{lcccc} \toprule
				& Share & Net  & \multicolumn{2}{c}{Optimal targeting vs.} \\
				& treated & donations & all-gift & no-gift \\
				\cline{2-5}    & (1)   & (2)   & (3)   & (4) \\
				\midrule		
			\multicolumn{5}{c}{\textbf{Panel A: Warm list}}\\\midrule
			
		Net donation amount & 0.334 & 17.35*** & 1.85***  & 1.79**	
 \\
		(1st year)		&       & (0.93) & (0.71)	& (0.72) 
\\
							\midrule		
			\multicolumn{5}{c}{\textbf{Panel B: Cold list}}\\\midrule
					Net donation amount & 0.014 & 0.15***&	0.99***&	-0.002
 \\
			(1st year)	&       & (0.02)&	(0.10)&	(0.013)
 \\
		\bottomrule
			\end{tabular}
		\end{center}
		\parbox{\textwidth}{\footnotesize{\textbf{Notes:} This table replicates the main results of Tables \ref{main_res_warm} and \ref{main_res_cold} when estimating all nuisance parameters with a cross-fitted lasso estimator. We do not penalize the confounders. We estimate the propensity score with a logit lasso. Standard errors are in parentheses. ***/**/* indicate statistical significance at the 1\%/5\%/10\% level. }}
	\end{table}

\end{appendix}
\end{document}